\documentclass[aps,prd,amssymb,nobibnotes, amsmath,twocolumn,nofootinbib,superscriptaddress,floatfix,preprintnumbers]{revtex4-1}

\usepackage{graphicx,color}% Include figure files
\usepackage{dcolumn}% Align table columns on decimal point
\usepackage{bm}% bold math
\usepackage{hyperref}% add hypertext capabilities
\usepackage{xurl}
\usepackage{amsmath}
\usepackage{mathtools}
\usepackage{multirow}
\usepackage{stackrel}
\usepackage[lofdepth,lotdepth,caption=false]{subfig}
\usepackage{tikz-feynman}
\usepackage{verbatim}
\usepackage{xcolor}

\usepackage{float}

\begin{document}

%===========================================================

\begin{flushleft}
LAPTH-032/26
\end{flushleft}

\title{Nuclei  in high-energy neutrino sources: A multimessenger study of in-source propagation}

\author{AmirFarzan Esmaeili}
\email{afesmaeili@unicamp.br}
\affiliation{Departamento de Raios Cósmicos e Cronologia, Instituto de F\'{\i}sica Gleb Wataghin, Universidade Estadual de Campinas, R. S\'ergio Buarque de Holanda, 777, Brazil}

\author{Arman Esmaili}
\email{arman@puc-rio.br}
\affiliation{Departamento de Física, Pontifícia Universidade Católica do Rio de Janeiro, Rio de Janeiro 22452-970, Brazil}

\author{Pasquale Dario~Serpico}
\email{serpico@lapth.cnrs.fr}
\affiliation{Laboratoire d’Annecy de Physique Théorique (LAPTh), CNRS, USMB, 74940 Annecy, France}

\begin{abstract}
The joint observation of astrophysical sources in gamma rays and neutrinos can provide invaluable insight into the physical conditions of the source, including its size, particle densities, and acceleration and production mechanisms. In this work, we investigate the role of nuclear composition in high-energy astrophysical environments. Using NGC 1068 as a representative example, we perform detailed Monte Carlo simulations of nuclear and electromagnetic cascades within the source and study the imprints of the injected nuclear composition on the resulting neutrino and gamma-ray emissions. We further discuss the importance of MeV–GeV gamma-ray observations for constraining the source composition in the context of future gamma-ray experiments. A dedicated re-analysis of archival COMPTEL observations is also presented. 
\end{abstract}
\maketitle

%%%%%%%%%%%%%%%%%%%%%%%%%%%%%%%%%%
%%%%%%%%%%%%%%%%%%%%%%%%%%%%%%%%%%
\section{Introduction}
%%%%%%%%%%%%%%%%%%%%%%%%%%%%%%%%%%
%%%%%%%%%%%%%%%%%%%%%%%%%%%%%%%%%%

The discovery of high-energy cosmic neutrinos by the IceCube Collaboration~\cite{IceCube:2013low,IceCube:2013cdw,IceCube:2015qii} has immediately raised numerous questions about their sources. The widespread pre-discovery expectation was that typical processes generating neutrinos, such as inelastic photohadronic ($p\gamma$) and hadronuclear ($pp$) interactions, should also produce gamma rays of similar energies, which could serve as diagnostic tools for uncovering the sources~\cite{Kelner:2006tc,Kamae:2006bf,1986A&A...157..223D}. To date, however, the quasi-totality of the IceCube sources remain without TeV or GeV gamma-ray counterparts. Even a cosmologically remote origin of these neutrinos, however, is insufficient to reconcile the phenomenology: while the absorption of TeV photons onto the extragalactic background light (EBL) would explain the lack of TeV counterparts, a significant contribution to the GeV diffuse gamma-ray background is expected, via electromagnetic cascading~\cite{Murase:2013rfa,Murase:2015xka}. Nonetheless, Fermi-LAT\footnote{\texttt{https://fermi.gsfc.nasa.gov/}} data tightly constrain this scenario, and the most widely accepted conclusion is that the majority of 10 - 100 TeV neutrino sources are ``hidden'', probably opaque to GeV-TeV gamma rays, see e.g.~\cite{Murase:2015xka,Capanema:2020rjj,Capanema:2020oet,Fang:2022trf}. 

The non-detection of gamma counterparts can also be used to constrain how compact the emission region is and the energy of the photon bath responsible for the opacity. Notably in the case
of the Seyfert galaxy NGC 1068, which is one of the few identified sources of neutrinos~\cite{IceCube:2022der}, these considerations translate into a tool to study non-thermal processes near the supermassive black hole (SMBH) at its center, as for instance in refs.~\cite{Das:2024vug,Das:2026rwt}.

Here, we want to extend these studies towards a less well-explored direction: the chemical composition of the accelerated charged particles which originate the neutrino flux in the first place. Are they dominantly protons, as almost always assumed, or do they contain a sizable fraction of nuclei? If yes, are we dealing with rather light nuclei as helium or heavy ones like iron? What if an ultra-heavy composition is relevant, as it has been invoked to better explain ultra high energy cosmic-ray (UHECR) data~\cite{Farrar:2024zsm,Zhang:2024sjp}? How relevant is the reprocessing of the chemical composition within the source? 

Similar questions have been raised and discussed in the UHECR context. In particular, photodisintegration of accelerated nuclei in the radiation field surrounding the accelerator can naturally reshape the escaping spectrum and composition, and has been used to model the ``ankle'' region and the associated neutrino signatures. Subsequent multimessenger studies have further explored how UHECR source environments can be constrained by cosmic-ray, neutrino, and gamma-ray data~\cite{Unger:2015laa, Muzio:2019leu, Muzio:2021zud}.
Our goal in this article is to apply an analogous approach to the neutrino sources, clarifying how relaxing the pure proton composition hypothesis alters the multimessenger phenomenology, such as the photon vs. neutrino yields and spectra or the overall luminosity requirements. One aspect of our article is methodological, with a new numerical treatment of nuclear-induced cascades. We also include some neglected processes, commenting on the level of accuracy of the approximation which is typically used (see Sec.~\ref{sec:nuc_cascades}). In Sec.~\ref{sec:multimessenger}, the non-trivial variation of yields and spectra with environmental parameters of the sources is explored in a simplified toy-model of a source.
As for the phenomenological applications, in Sec.~\ref{sec:model} we will focus on the NGC 1068 case and, to ease comparison with the existing literature, we put ourselves in the same setup of ref.~\cite{Das:2024vug}. Interestingly, even current MeV data turn out to be (mildly) constraining, which motivates us to perform a dedicated re-analysis of the Imaging Compton Telescope-COMPTEL\footnote{\texttt{https://heasarc.gsfc.nasa.gov/docs/cgro/comptel/}} archival data (Sec.~\ref{subsec:comptel_ngc1068}). A discussion of our results and our conclusions are in Sec.~\ref{sec:conclusion}.

%%%%%%%%%%%%%%%%%%%%%%%%%%%%%%%%%%
%%%%%%%%%%%%%%%%%%%%%%%%%%%%%%%%%%
\section{Electromagnetic processes: physical and numerical aspects}\label{sec:nuc_cascades}
%%%%%%%%%%%%%%%%%%%%%%%%%%%%%%%%%%
%%%%%%%%%%%%%%%%%%%%%%%%%%%%%%%%%%

The principal framework for explaining the bulk of neutrino emission from an astrophysical site observed by IceCube consists of the \textit{photomeson} (PM) interactions of accelerated  nucleons or nuclei in a medium accommodating sufficiently energetic photons.

The neutrino yield is mostly associated with charged pion decays and subsequent muon decays; if the medium were optically thin, neutral pion decays would generate gamma-rays whose flux and energies are comparable with the neutrino one. This flux in the TeV sky is not seen, which could, in principle, be explained by cosmologically distant sources whose photons are absorbed by the EBL. However, the lack of a sizable diffuse gamma-ray component in the Fermi-LAT GeV data suggests that the medium surrounding the acceleration site is opaque: one must thus consider further \textit{absorption processes} affecting the $\pi^0$-produced photons, the $e^+$ and $e^-$ byproducts of the charged pion decays, as well as the $e^\pm$ pairs (\textit{Bethe-Heitler or BH process}) unavoidably associated with the primaries propagating in the energetic photon bath. 
For the specific case of nuclear (as opposed to nucleon) primaries, one should further include at least the \textit{photodisintegration process} to evolve the primary nucleus energy and chemical composition. 
The discussion above motivates us to model the in-source propagation of nuclei, and the associated neutrino and electromagnetic emissions, by a Monte Carlo simulation performed in two stages: \textit{i}) photonuclear interactions and nuclear cascade evolution; \textit{ii}) electromagnetic cascade development. 

In stage (\textit{i}), as described in the following, in addition to including all the processes considered in the existing studies \cite{Boncioli:2016lkt, Biehl:2017zlw, Zhang:2023ewt}, we extend the photonuclear processes up to $Z=90$ isotopes and we account for the first time for the  pair production with the capture in the $L$ shell, following our previous study~\cite{Esmaeili:2024pwh}. Although the latter is subleading in energy-loss processes, we assess its quantitative role here.

In stage (\textit{ii}), photons and electrons/positrons produced in stage (\textit{i}) are propagated through the assumed radiation field surrounding the source. Their electron pair production (EPP) and inverse Compton (ICS) interactions drive a cascade, degrading and redistributing the injected energy before it emerges as the observable gamma-ray signal. Technically, we also include more rare processes, namely double pair production (DPP) and muon pair production (MPP) for photon--photon interactions, and electron triplet production (ETP) and electron--muon pair production (EMPP) for electron--photon interactions, as we detailed in refs.~\cite{Esmaeili:2022cpz,Esmaeili:2023vyk}; however they are negligible in the dynamical range covered here. On the other hand, we neglect synchrotron losses: qualitatively, they would shift further the electromagnetic cascade to lower energies. Their inclusion is left for future work.

More in detail, the source is modeled as homogeneous region of size $R$. At each step, the Monte Carlo samples the free-flight distance of the projectile conditionally on the remaining distance to the source boundary: the total rate of the enabled discrete processes determines the probability that an interaction occurs before escape, while their relative rates determine which channel is realized at the interaction point. Continuous loss processes are considered along the sampled propagation segment, so that energy losses and the associated secondary injection are tied to the actual distance traveled inside the source. In the remaining of this section, we are going to describe the each process entering stage (\textit{i}) separately. For details on our treatment of stage (\textit{ii}), see refs.~\cite{Esmaeili:2022cpz,Esmaeili:2023vyk}.

%%%%%%%%%%%%%%%%%%%%%%%%%%%%
%%%%%%%%%%%%%%%%%%%%%%%%%%%%
\subsubsection{Photodisintegration (PD)}
%%%%%%%%%%%%%%%%%%%%%%%%%%%%
%%%%%%%%%%%%%%%%%%%%%%%%%%%%

A nucleus of mass number $A$ and charge $Z$ can be excited by absorbing a background photon in its rest frame and de-excite by ejecting one or more photons, nucleons, or light nuclear fragments. This changes the nuclear composition during propagation and produces secondary fragments that may continue interacting in the source. 

In our Monte Carlo simulation, photodisintegration (PD) is implemented as an exclusive-channel process. Each breakup channel is treated as a separate discrete interaction, with its own interaction length computed in the source photon field. The enabled channels compete with all other discrete nuclear processes through their corresponding rates. Once photodisintegration is selected at the interaction vertex, the specific channel determines the change in nuclear mass and charge, $(A,Z)\rightarrow(A-\Delta A,Z-\Delta Z)$, as well as the light fragments and de-excitation photons injected into the secondary particle list, where the latter contributes to the electromagnetic cascade development.

For intermediate and heavy nuclei, we use photodisintegration cross sections based on TALYS/TENDL-2023 tables \cite{Koning:2019qbo, Koning:2023ixl}\footnote{\url{https://tendl.imperial.ac.uk/}}\footnote{For recent analytic fitting formulae for photodisintegration cross sections of heavy and ultraheavy nuclei, see \cite{Ekanger:2026zqd}.}. The tabulated network contains 1989 isotopes, covering $4\leq Z\leq 90$ and $10\leq A\leq 239$. For each isotope, the cross sections are provided as functions of the photon energy in the nuclear rest frame and include total photodisintegration as well as exclusive partial channels including the so-called nuclear resonance fluorescence (NRF) $(\gamma,\gamma)$, one- and multi-nucleon emissions $(\gamma,n)$, $(\gamma,2n)$, $(\gamma,3n)$, $(\gamma,4n)$, $(\gamma,p)$, $(\gamma,2p)$, $(\gamma,3p)$; and light-fragment emissions $(\gamma,np)$, $(\gamma,d)$, $(\gamma,t)$, $(\gamma,{}^3{\rm He})$, $(\gamma,\alpha)$, where $n$, $p$, $d$ and $t$ are denoting the neutron, proton, deuteron and tritium.

For light nuclei, we use a dedicated photodisintegration network. For $A=2,3,4$, the total cross sections and the relevant breakup channels are described using the parametrizations of \cite{Soriano:2018lly}. For these nuclei, the code separates the three-body $np$ breakup from the remaining two-body channels. In particular, for ${}^4{\rm He}$ the $np$ channel is described by a Bethe--Peierls form, while the remaining strength is divided between the one-neutron and one-proton channels. For ${}^3{\rm He}$ and $t$, the corresponding two-body breakup is assigned to the physically allowed nucleon-emission channel. For ${}^9{\rm Be}$, we use the TENDL table of ${}^9{\rm Be}(\gamma,n){}^8{\rm Be}$ channel. For ${}^6{\rm Li}$, ${}^7{\rm Li}$, and ${}^7{\rm Be}$, the one-nucleon channels are modeled using the giant dipole resonance (GDR) component of the Kossov parametrization \cite{Kossov:2002}, while the cluster $\alpha$ channels of ${}^7{\rm Li}$ and ${}^7{\rm Be}$ are included using the fits of ref.~\cite{Ishida:2014wqa}. Short-lived light nuclei produced during the cascade are not propagated as independent species: ${}^5{\rm He}$, ${}^5{\rm Li}$, ${}^6{\rm Be}$, ${}^7{\rm He}$, and ${}^8{\rm Be}$ are immediately mapped onto their prompt fragmentation channels, while ${}^6{\rm He}$ is mapped to ${}^6{\rm Li}$ in the reduced network. The following prescription closes the light-element sector used in the propagation:
\[
{}^5{\rm He}\rightarrow{}^4{\rm He}+n,\qquad
{}^5{\rm Li}\rightarrow{}^4{\rm He}+p,\qquad
{}^6{\rm Be}\rightarrow{}^4{\rm He}+2p,
\]
\[
{}^7{\rm He}\rightarrow{}^4{\rm He}+3n,\qquad
{}^8{\rm Be}\rightarrow 2\alpha,\qquad
{}^6{\rm He}\rightarrow{}^6{\rm Li}.
\]

At a baryon-removal photodisintegration vertex, the energy of the parent nucleus is partitioned between the heavy residual and the emitted fragment(s). We use the high-energy approximation in which the fraction of the projectile energy carried by the emitted fragments is
\begin{equation}
\eta = \frac{\Delta A}{A}.
\end{equation}
The residual nucleus therefore carries a fraction $1-\eta$ of the incoming energy, while the emitted light fragments share the remaining energy according to the selected exclusive channel. This prescription preserves the leading energy flow of the nuclear cascade and is adequate in the ultra-relativistic regime considered here, where nuclear binding and recoil energies are small compared to the projectile energy.

The same baryon-removal events are accompanied by prompt nuclear de-excitation emission. We model this component by injecting $N_{\gamma}^{\rm deex}=3$ photons with rest-frame energy $E_{\gamma}^{\prime}=2\,{\rm MeV}$ per event \cite{Zhang:2023ewt, Murase:2010va}. In the source frame, each photon energy is boosted by the Lorentz factor of the residual nucleus, $E_{\gamma}\simeq \Gamma_A E_{\gamma}^{\prime}$, and the corresponding energy is subtracted from the residual. These photons are then propagated as part of the electromagnetic cascade. We denote these photons by PD de-ex through out this paper.

The $(\gamma,\gamma)$ channel (NRF) is treated separately. In this case the nuclear species is unchanged, but gamma-ray line photons are emitted with a  multiplicity taken from the TENDL-2023 gamma-production tables associated with the corresponding isotope. These photons are stored with a distinct origin tag and are propagated through the electromagnetic cascade in the same way as the other photon secondaries.
%%%%%%%%%%%%%%%%%%%%%%%%
%%%%%%%%%%%%%%%%%%%%%%%%
\subsubsection{Photomeson (PM)}
%%%%%%%%%%%%%%%%%%%%%%%%
%%%%%%%%%%%%%%%%%%%%%%%%

Photomeson (PM) production is included as a discrete photonuclear channel. Its interaction length is computed using the photonuclear cross section parametrization of Kossov \cite{Kossov:2002}. For nucleons, we use the corresponding nucleon fit. For nuclei, we use the $A$-dependent above-GDR part of the parametrization, retaining the quasi-deuteron, $\Delta$-resonance, transition/Roper-like, and Regge--Pomeron contributions. The giant dipole resonance (GDR) contribution is excluded from this channel, since it is already accounted for in the photodisintegration treatment.

Once a photomeson interaction is selected by the Monte Carlo, the final-state particles are generated with SOPHIA~\cite{Mucke:1999yb}\footnote{\href{https://www.uibk.ac.at/projects/he-cosmic-sources/tools/sophia/index.html.en}{SOPHIA website}}, using its modern SOPHIA-next implementation. For nuclei, we adopt a superposition approximation: the interacting nucleon is selected as a proton with the probability $Z/A$ and as a neutron with the probability $1-Z/A$, and the event generator is called with the corresponding per-nucleon energy, $E_A/A$, where $E_A$ denotes the total energy of the injected nucleus of mass number $A$. 

The outgoing leading baryon from the SOPHIA event is used to update the transported nuclear remnant. In particular, charge-exchange reactions are included by changing the nuclear charge according to $p\rightarrow n$ or $n\rightarrow p$, while keeping the mass number fixed in the spectator approximation. The energy of the remnant is updated by replacing the energy of the interacting nucleon by that of the leading outgoing baryon, whereas the remaining spectator nucleons are kept implicit. Photons and electrons/positrons produced in the event are injected into the electromagnetic cascade, and neutrinos are recorded as part of the source neutrino output.

%%%%%%%%%%%%%%%%%%%%%%%%
%%%%%%%%%%%%%%%%%%%%%%%%
\subsubsection{Bethe-Heitler (BH) Pair Production}
%%%%%%%%%%%%%%%%%%%%%%%%
%%%%%%%%%%%%%%%%%%%%%%%%

Bethe--Heitler (BH) pair production,
\begin{equation}
A_Z+\gamma \rightarrow A_Z + e^+ + e^-~,
\end{equation}
is included as a continuous energy-loss process for charged nuclei, using the cross section based on the parametrization of ref.~\cite{1992ApJ...400..181C}. The nuclear species is left unchanged, while the produced $e^\pm$ pairs are injected into the electromagnetic cascade.

In each propagation segment, the produced electrons and positrons are sampled from the Bethe--Heitler lepton spectrum \cite{Kelner:2008ke}, and their total energy is subtracted from the projectile nucleus. The process is therefore continuous at the level of the leading nuclear trajectory, while the associated $e^\pm$ secondaries are retained explicitly and injected into the electromagnetic cascade.

For numerical stability in optically thick regions, the continuous-loss update is adaptively sub-stepped whenever the fractional energy loss over a segment exceeds a prescribed threshold. In the runs presented here, this threshold is set to $\Delta E/E_{A}=10^{-2}$.

%%%%%%%%%%%%%%%%%%%%%%%
%%%%%%%%%%%%%%%%%%%%%%%
\subsubsection{Pair Production with Capture in the L shell}
%%%%%%%%%%%%%%%%%%%%%%%
%%%%%%%%%%%%%%%%%%%%%%%
Pair production with capture (PPC) is an analogue of the Bethe--Heitler pair production, with the final state $e^-$ captured directly into a bound atomic state with quantum numbers $n\ell j$,
\begin{equation}
A_Z+\gamma \rightarrow A_Z(e^-_{n\ell j}) + e^+\,.
\end{equation}

PPC has been discussed in the context of charge-state evolution of ultra-high-energy nuclei in photon backgrounds, in particular for capture into the $K$ shell \cite{Esmaeili:2024pwh}. Here we also include the capture into the $L$ shell, which provides an additional source of line photons after the captured electron makes a transition to the $K$ shell.

The PPC cross sections used in this work are based on the bound-free pair production calculations of ref.~\cite{Agger:1997ps}, where Coulomb wave functions have been used to compute the photon-impact pair production with capture into both $K$ and $L$ shells. In the Monte Carlo, capture into the $L$ shell (hereafter PPCL) is implemented through three independent subshell channels: $2s$, $2p_{1/2}$, and $2p_{3/2}$. Each subshell has its own interaction length in the source photon field and enters the same discrete-channel competition as photodisintegration and photomeson production.

After $L$-shell capture, the bound electron is assumed to de-excite promptly to the $K$ shell. The associated atomic transition produces a line photon with rest-frame energy
\begin{equation}
\varepsilon_{\gamma,L\rightarrow K}=\omega_K(Z)-\omega_L^{(i)}(Z)~,
\end{equation}
where $\omega_K$ is the $K$ shell binding energy and $\omega_L^{(i)}$ denotes the binding energy of the selected $L$ subshell, $i\in\{2s,2p_{1/2},2p_{3/2}\}$. In the source frame, the emitted photon energy is boosted by the Lorentz factor of the nucleus $\Gamma_A$, as
\begin{equation}\label{eq:ppcl_line_energy}
E_{\gamma,L\rightarrow K}\simeq\Gamma_A\, \varepsilon_{\gamma,L\rightarrow K}\,.
\end{equation}
These photons are stored with a dedicated PPCL origin tag and propagated through the electromagnetic cascade together with the other photon secondaries.

In the present implementation, the nuclear charge is left unchanged in PPCL events, and the captured-electron charge state is not evolved as an independent degree of freedom. The process is included only through its contribution to the electromagnetic cascade, via the prompt line photon emitted in the subsequent atomic de-excitation. 

One notice is in order: The energy injected by the accompanying positron into the electromagnetic cascade can be safely neglected. In the nucleus rest frame, its energy is fixed by energy conservation,
\begin{equation}
\varepsilon_{e^+}=\varepsilon_{\gamma_b}-m_e+\omega_L^{(i)}(Z),
\end{equation}
where $\varepsilon_{e^+}$ and $\varepsilon_{\gamma_b}$ are the positron and background-photon energies in that frame, respectively. The differential bound-free pair-production cross section is peaked for positron emission along the incident photon direction, i.e. for $\theta_{\gamma e^+}\simeq 0$, where $\theta_{\gamma e^+}$ is the angle between the outgoing positron and the incoming photon in the nucleus rest frame (See ref.~\cite{Agger:1997ps}). Since PPCL interactions of ultra-relativistic nuclei are dominated by nearly head-on target photons, this corresponds approximately to emission opposite to the nuclear boost direction. The lab frame positron energy, $E_{e^+}$, is therefore de-boosted,
\begin{equation}
E_{e^+}
\simeq
\Gamma_A\left(\varepsilon_{e^+}-\beta_A p_{e^+}\right)
\simeq
\frac{\varepsilon_{e^+}}{2\Gamma_A},
\end{equation}
where $\beta_A$ is the nuclear velocity and $p_{e^+}=\sqrt{\varepsilon_{e^+}^2-m_e^2}$ is the positron momentum in the nucleus rest frame. In the last step we assumed a relativistic positron, $p_{e^+}\simeq \varepsilon_{e^+}$, and $\Gamma_A\gg1$. Comparing with the line photon energy in Eq.~\eqref{eq:ppcl_line_energy},
\begin{equation}
\frac{E_{e^+}}{E_{\gamma,L\rightarrow K}}
\simeq
\frac{1}{2\Gamma_A^2}\frac{\varepsilon_{e^+}}{ \varepsilon_{\gamma,L\rightarrow K}}
\ll 1 .
\end{equation}
We therefore neglect direct positron injection from PPCL and retain only the prompt $L\rightarrow K$ line photon as the electromagnetic secondary associated with this channel.

%%%%%%%%%%%%%%%%%%%%
%%%%%%%%%%%%%%%%%%%%
\subsubsection{Nuclear Decay}
%%%%%%%%%%%%%%%%%%%%
%%%%%%%%%%%%%%%%%%%%

Nuclear beta decays are included in the propagation menu for completeness. The decay lengths are computed from tabulated rest-frame partial beta-decay rates, derived from the IAEA LiveChart database \cite{IAEA_LiveChart} for the same set of 1989 isotopes used in the TALYS/TENDL photodisintegration network. Since the propagated nuclei are bare, electron-capture channels are not included. Also, the bound-state beta decay is neglected.

For the source sizes and projectile energies considered in this work, the boosted decay lengths are much larger than both the escape scale and the relevant interaction lengths. Therefore, nuclear beta decay has a negligible impact on the in-source evolution of the composition and on the electromagnetic and neutrino outputs. In particular, this applies to neutrons that, under the conditions of interest, are basically stable over the neutrino-production region: their decay length is $\ell_{\rm dec}=\Gamma\,c \tau_{\rm dec}\sim 3\times 10^{13}\Gamma\,$cm, much larger than the source size considered below, once accounting that the relevant boost factor are $10^3\lesssim \Gamma\lesssim 10^7$.

%%%%%%%%%%%%%%%%%%%%
%%%%%%%%%%%%%%%%%%%%
\section{Multimessenger Discussion}\label{sec:multimessenger}
%%%%%%%%%%%%%%%%%%%%
%%%%%%%%%%%%%%%%%%%%
The ratio of neutrino to electromagnetic yield for nuclear vs. proton injection is a non-trivial (for instance, non-monotonic) function of the (astro)physical scales in the problem. Since the relevant phenomenology has not been elucidated before, we
start by illustrating the trends in a highly idealised configuration, capturing the essential ingredients of the astrophysical system we have in mind. If we consider a SMBH of mass $10^7 M_{\odot}$, its benchmark spatial scale has size $R=R_S\equiv 2GM/c^2=2.95\times 10^{12}\,$cm, i.e. the SMBH Schwarzschild radius\footnote{More precisely, we imagine that production takes place in a homogeneous region of space of size $R$ which, while close to the SMBH,  is sufficiently far for general relativistic effects to be qualitatively irrelevant in altering the phenomenology.}. SMBHs are surrounded by a disk, with a dense UV photon spectrum, and a hotter but more rarefied corona.  In this section, we simplify the disk–corona radiation field by the superposition of two broadened quasi-monochromatic components: \textit{i}) A UV disk component, centered at $\epsilon_{\rm d}=20\,{\rm eV}$, with Gaussian width $\sigma_{\rm d}=\epsilon_{\rm d}$ and number density $n_{\gamma,{\rm d}}$. \textit{ii}) An X-ray coronal component, centered at $\epsilon_{\rm c}=2\,{\rm keV}$, with Gaussian width $\sigma_{\rm c}=\epsilon_{\rm c}$, with its number density fixed to be a small fraction of the disk component,
$n_{\gamma,{\rm c}}=3\times10^{-3}\,n_{\gamma,{\rm d}}$.
The total photon density is therefore
$n_\gamma=n_{\gamma,{\rm d}}+n_{\gamma,{\rm c}}
\simeq n_{\gamma,{\rm d}}$.  The multimessenger yields are controlled by the optical depths of the astrophysical site for the relevant microphysics processes, responsible for reprocessing the injected nuclei before escape. For a fixed source size $R$, as considered in this section, the optical depths are set by the photon density $n_\gamma$. 
We consider three benchmark values of this total density,
$n_\gamma=3\times10^{16},\;3\times10^{17},\;3\times10^{18}\,{\rm cm^{-3}}$, covering rarefied to dense photon fields.
The relative importance of the competing interaction channels determines how the injected energy is partitioned among neutrinos, escaping nuclear fragments, and electromagnetic secondaries.

%%%%%%%%%%%%%%%%%%%%%%%%%%%%%%%
%%%%%%%%%%%% FIG. 1 %%%%%%%%%%%
\begin{figure}[t!]
  \centering
  \includegraphics[width=0.99\columnwidth]{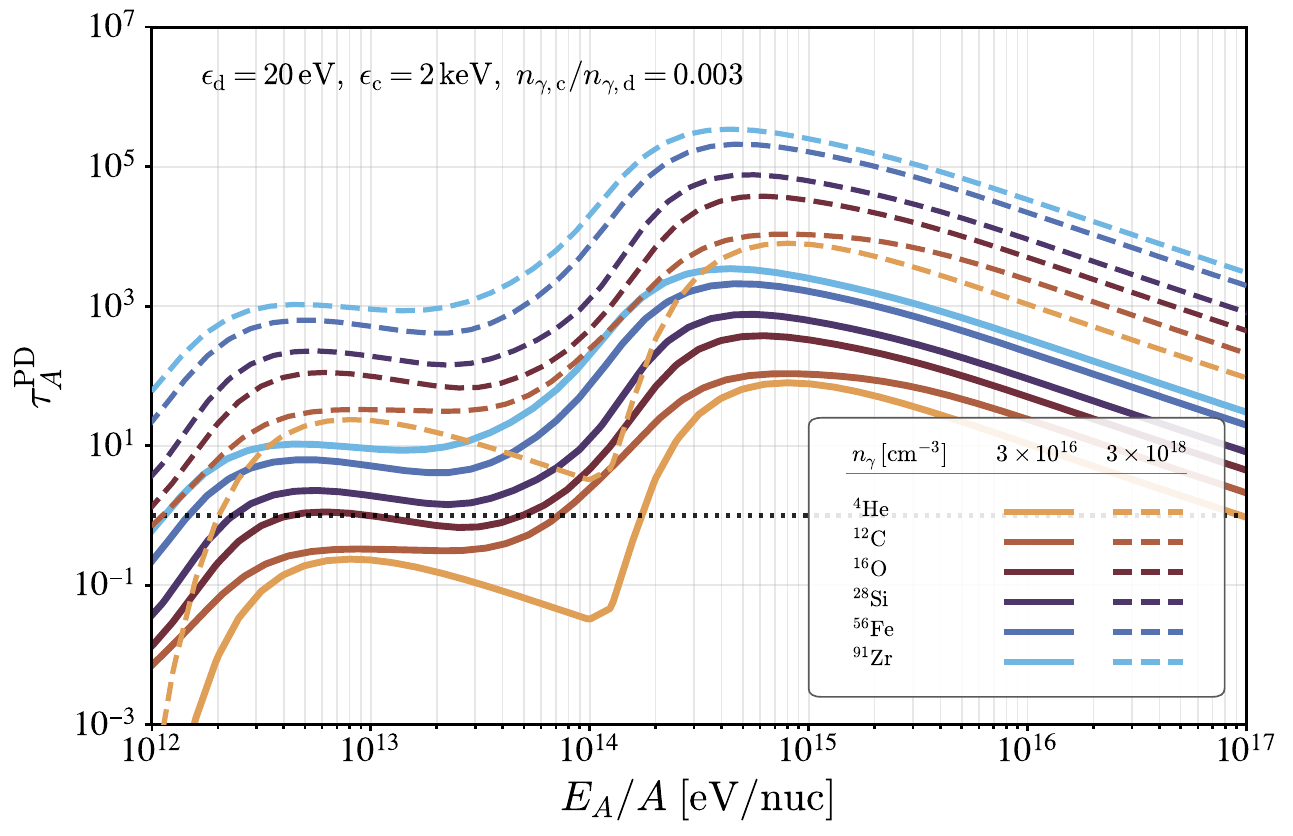}
  \caption{Photodisintegration optical depth, $\tau^{\rm PD}_A=R/\lambda^{\rm PD}_A$, for representative nuclei propagating in the simplified model described in the text. The curves are shown as a function of energy per nucleon, $E_A/A$ , for a low (solid lines) and a high (dashed lines) density of target phoyons, and for different nuclear primaries (color coded). }
  \label{fig:pd_optical_depth_doublegauss}
\end{figure}
%%%%%%%%%%%%%%%%%%%%%%%%%%%%%%%
%%%%%%%%%%%%%%%%%%%%%%%%%%%%%%%

Figure~\ref{fig:pd_optical_depth_doublegauss} shows the photodisintegration (PD) optical depth $\tau_A^{\rm PD}\equiv R/\lambda_A^{\rm PD}$, for several nuclei as a function of energy per nucleon, $E_A/A$. Here, $\lambda_A^{\rm PD}$ is the photodisintegration interaction length of the nucleus $A$. The different curves correspond to different nuclear species, color-coded in the figure. We show $\tau_A^{\rm PD}$ for two representative photon densities: a relatively rarefied source, $n_\gamma=3\times 10^{16}~{\rm cm^{-3}}$, and a denser source, $n_\gamma=3\times 10^{18}~{\rm cm^{-3}}$. The curves have similar shapes, reflecting the common Lorentz factor scaling of the interaction kinematics. The overall opacity scale, however, depends strongly on both $n_\gamma$ and the mass number $A$ of the primary. For the higher density, at all mass numbers and energies, the system is very  optically thick to PD: all nuclei get in fact destroyed into their constituents nucleons when traveling over distances well below $R$. For the lower density, this is only true for heavier nuclei and/or large energies: we expect in general a sizable amount of nucleons to survive locked into nuclei over a distance $R$. A first element to take into account is thus the degree to which the injected nucleus is disintegrated before escaping the source.

\begin{figure}[t!]
  \centering
  \includegraphics[width=0.99\columnwidth]{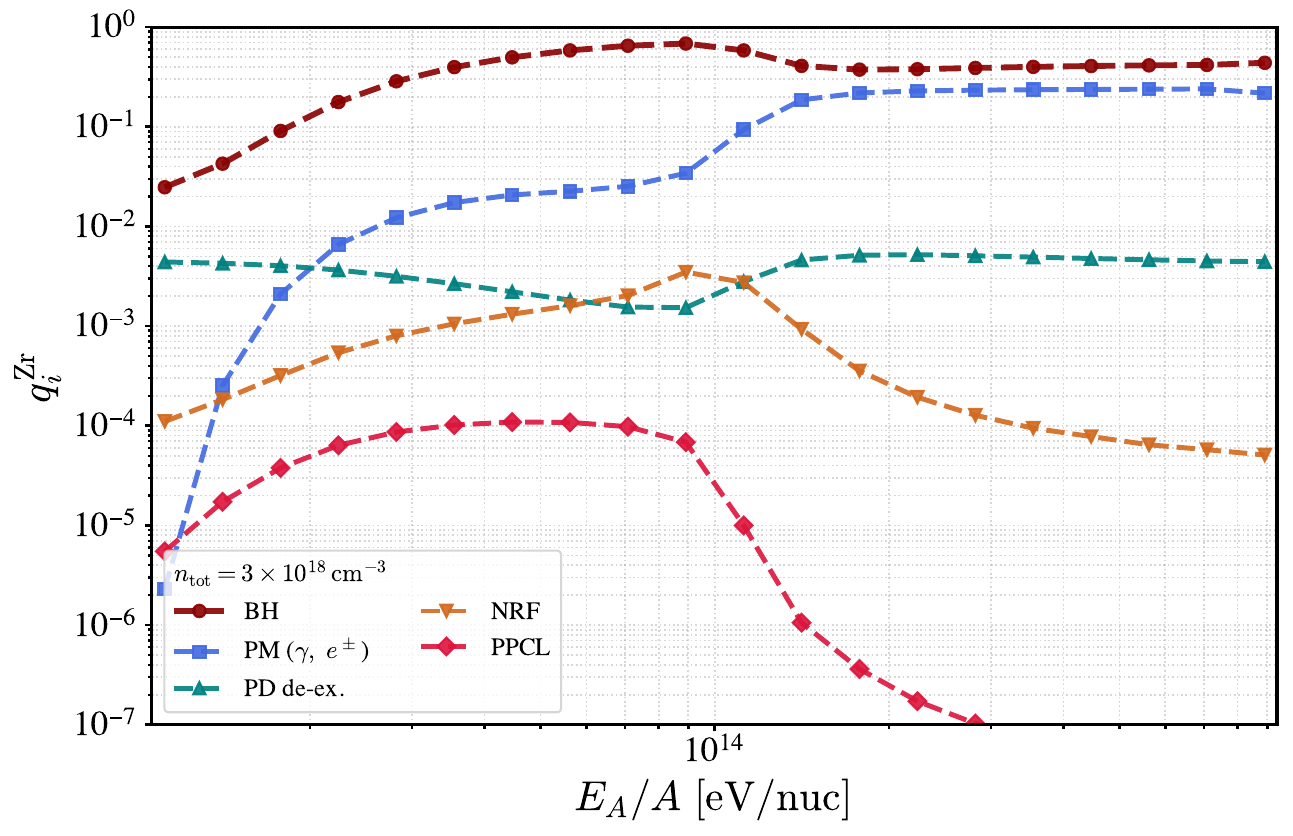}
  \caption{Energy fraction transferred to electromagnetic secondary channels for $^{91}{\rm Zr}$ injection in the double-Gaussian photon field, shown as a function of energy per nucleon $E_A/A$. The benchmark shown corresponds to $n_{\rm tot}=3\times10^{18}\,{\rm cm}^{-3}$.
  }
  \label{fig:zr_origin_budget}
\end{figure}

A second point concerns the energy losses. PD preserves the overall energy carried by nucleons; more specifically, the energy per nucleon is roughly conserved in this process. PPM production, when operative, degrades the energy/nucleon of the primary.  However, at fixed $E_A/A$, it scales linearly with $A$, so that it does not distinguish between different nuclear species, once the comparison is made at the same energy per nucleon. Species-dependent energy losses are therefore largely controlled by BH pair production, whose fractional loss rate scales as $Z^2/A$. Defining the effective BH optical depths as $\tau_p^{\rm BH}\equiv R/\ell_p^{\rm BH}$ and $\tau_A^{\rm BH}\equiv R/\ell_A^{\rm BH}$, where $\ell_{p}^{\rm BH}$ and $\ell_{A}^{\rm BH}$ are respectively the BH energy loss lengths for protons and nuclei of mass number $A$, one has $\tau_A^{\rm BH}>\tau_p^{\rm BH}$ for heavy nuclei. This loss channel vanishes for neutrons. The species-dependent behaviour therefore relies on a subtle interplay of PD and BH losses: 
Efficient photodisintegration intervening before significant BH losses produces a nucleon-dominated population inside the source, including free neutrons which are not subject to the electromagnetic losses compared to charged particles. On the other hand, when a significant fraction of the injected energy remains locked in heavy fragments, they experience losses larger than a population of protons of the same energy per nucleon, and their electromagnetic yield compared to the neutrino one is enhanced. 
To quantify the relative importance of energy losses, we introduce ``fractional energy'' functions $q^X_i(E)$ that describe, for a primary nucleus of type $X$ and energy $E$, the fraction of its energy channeled into the channel $i$. Figure~\ref{fig:zr_origin_budget} illustrates the  case of zirconium ($^{91}{\rm Zr}$) injection in the densest setup, $n_\gamma=3\times10^{18}\,{\rm cm}^{-3}$, for $E/A$ around $10^{14}\,$eV/nuc. The energy budget is dominated by BH pair production over all of the energy range. PM secondaries provide the next-largest contribution, reaching the $\sim 20\%$ level once the nucleon cascade enters the PM-thick regime. By contrast, the prompt nuclear de-excitation component remains subleading, below the percent level. The line-like channels are even smaller: the NRF contribution is generally below $10^{-3}$, although it can reach a peak value of $q_{\rm NRF}\simeq 3\times10^{-3}$, while PPCL never exceeds $q_{\rm PPCL}\sim10^{-4}$.

%%%%%%%%%%%%%%%%%%%%%%%%%%%%%%%
%%%%%%%%%%%% FIG. 4 %%%%%%%%%%%
%%%%%%%%%%%%%%%%%%%%%%%%%%%%%%% 

We also monitor the composition of the escaping nuclear components through the free-nucleon fractions
\begin{equation}
f_p\equiv
\frac{N_p^{\rm esc}}{A\,N_A^{\rm inj}}~,
\quad
f_n\equiv
\frac{N_n^{\rm esc}}{A\,N_A^{\rm inj}}~, \quad
    f_{p+n}\equiv
f_p+f_n~,
\end{equation}
where $N_A^{\rm inj}$ is the number of injected nuclei of mass number $A$ in each energy bin, and $N_p^{\rm esc}$ and $N_n^{\rm esc}$ are the numbers of free protons and neutrons  at the exit point of the region. 

\begin{figure*}[t!]
  \centering

  %----------------------------------------
  % Row 1: n_gamma = 3e16
  %----------------------------------------
  \subfloat[]{
    \begin{minipage}{0.485\textwidth}
      \centering
      \includegraphics[width=\linewidth]{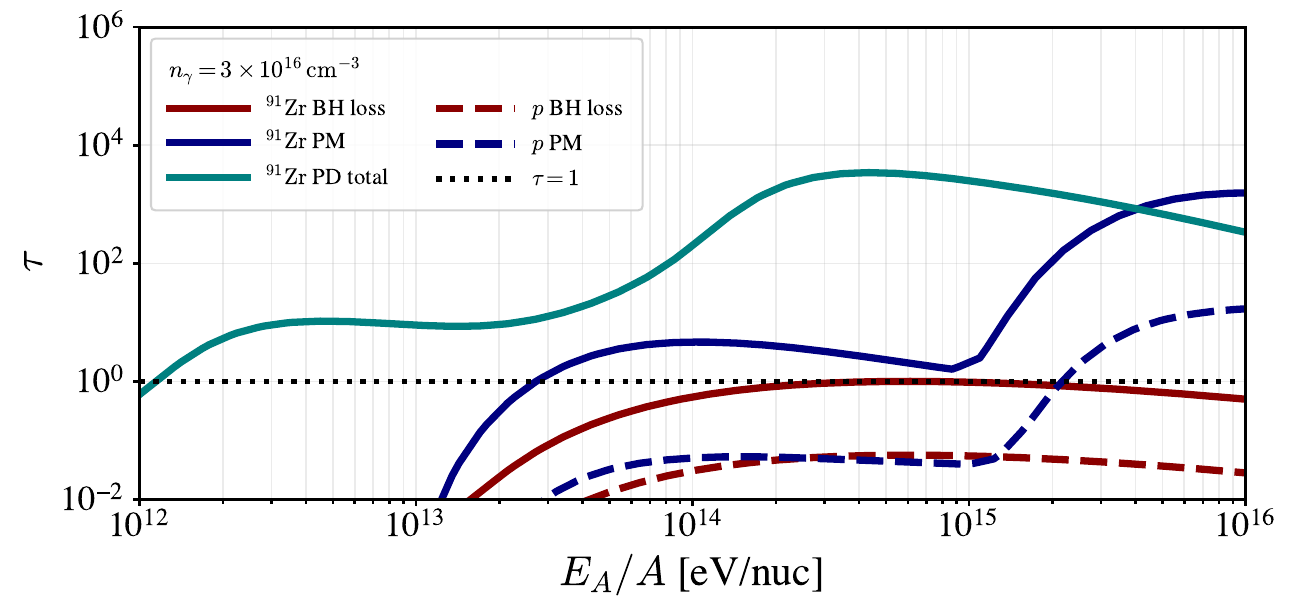}
      \label{fig:opticaldepths_zr_p_doublegauss_3e16}
    \end{minipage}
  }
  \hfill
  \subfloat[]{
    \begin{minipage}{0.485\textwidth}
      \centering
      \includegraphics[width=\linewidth]{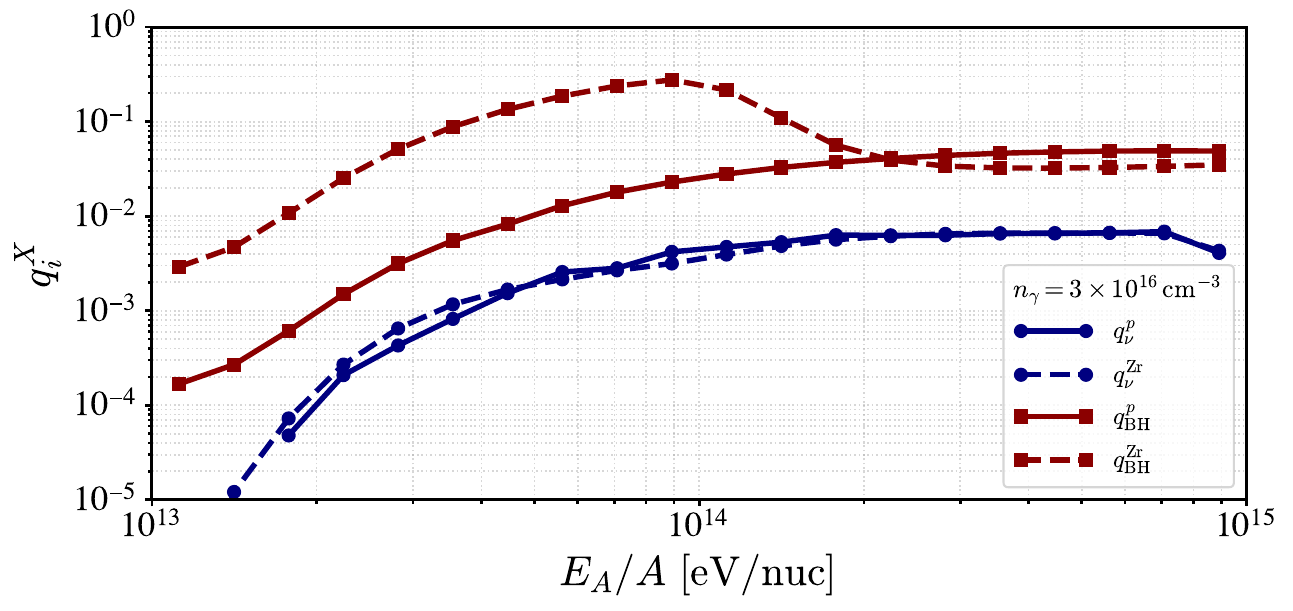}
      \label{fig:Zr_p_binbybin_budgets_doublegauss_ngtot3e16}
    \end{minipage}
  }

  \vspace{0.35cm}

  %----------------------------------------
  % Row 2: n_gamma = 3e17
  %----------------------------------------
  \subfloat[]{
    \begin{minipage}{0.485\textwidth}
      \centering
      \includegraphics[width=\linewidth]{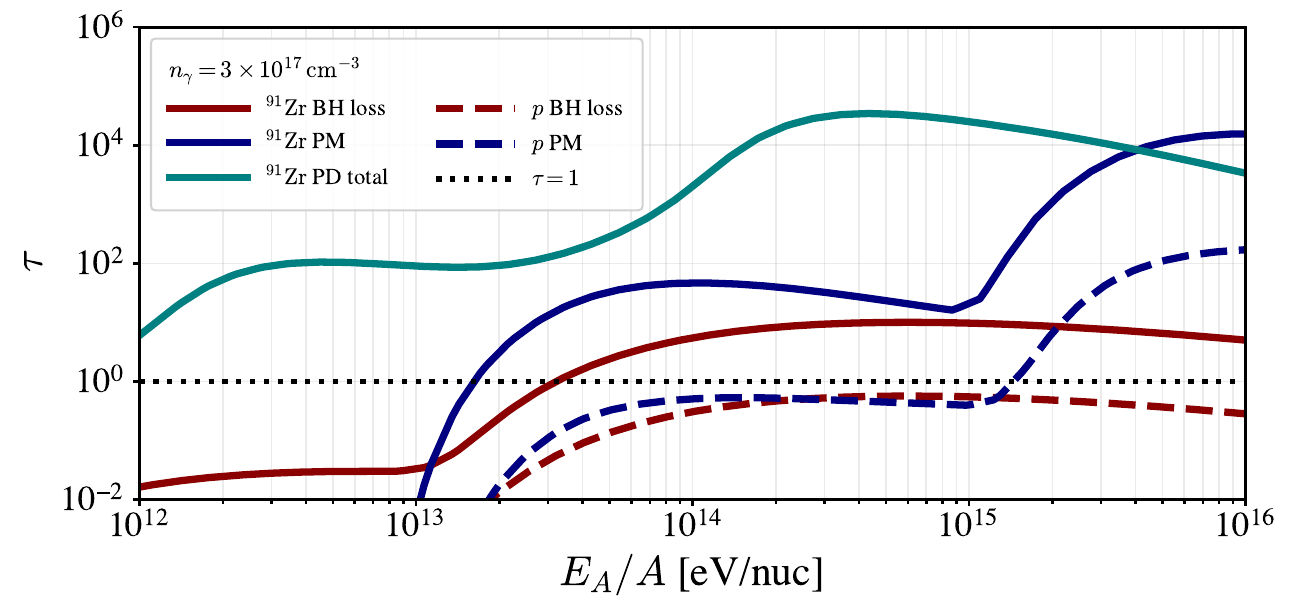}
      \label{fig:opticaldepths_zr_p_doublegauss_3e17}
    \end{minipage}
  }
  \hfill
  \subfloat[]{
    \begin{minipage}{0.485\textwidth}
      \centering
      \includegraphics[width=\linewidth]{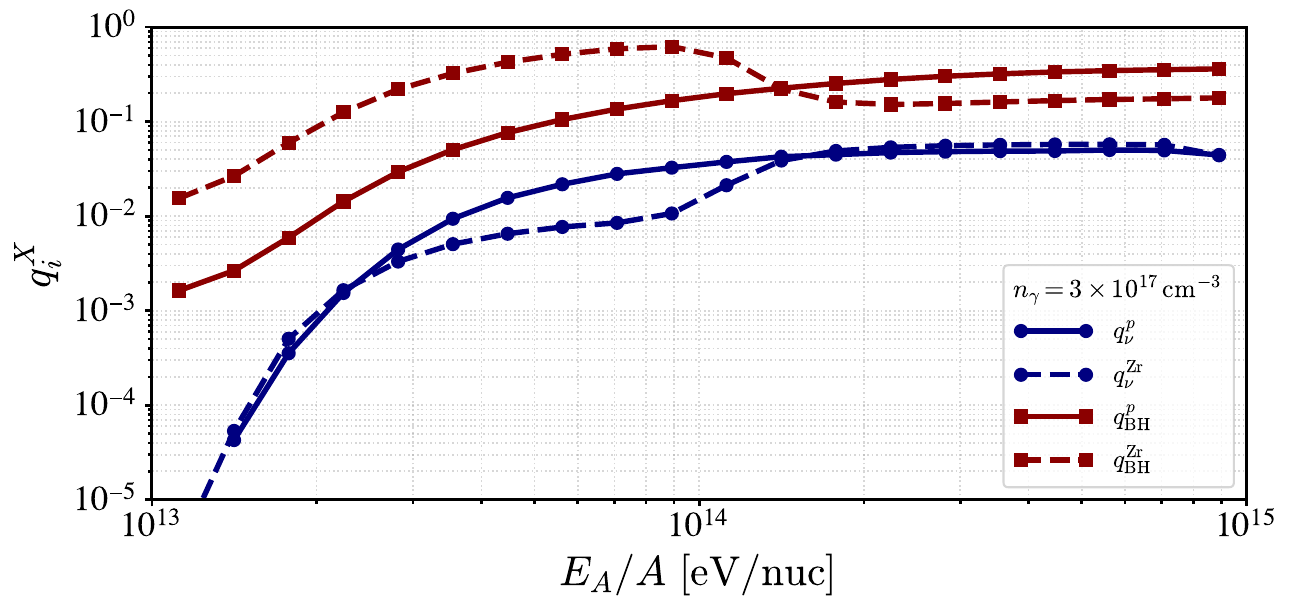}
      \label{fig:Zr_p_binbybin_budgets_doublegauss_ngtot3e17}
    \end{minipage}
  }

  \vspace{0.35cm}

  %----------------------------------------
  % Row 3: n_gamma = 3e18
  %----------------------------------------
  \subfloat[]{
    \begin{minipage}{0.485\textwidth}
      \centering
      \includegraphics[width=\linewidth]{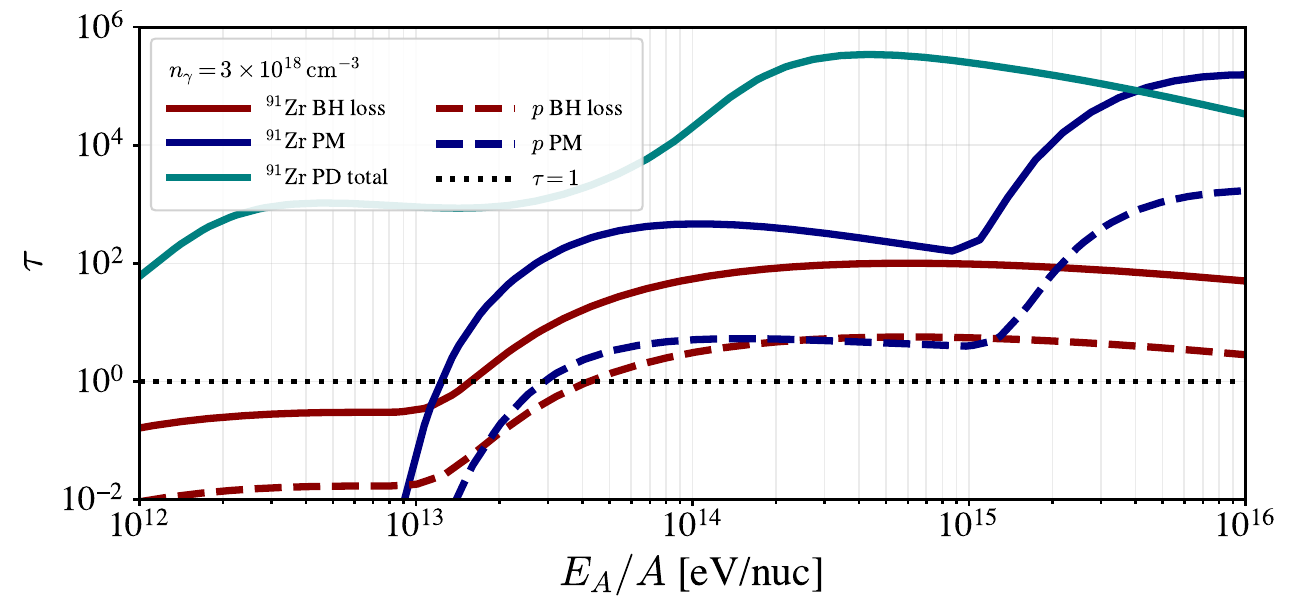}
      \label{fig:opticaldepths_zr_p_doublegauss_3e18}
    \end{minipage}
  }
  \hfill
  \subfloat[]{
    \begin{minipage}{0.485\textwidth}
      \centering
      \includegraphics[width=\linewidth]{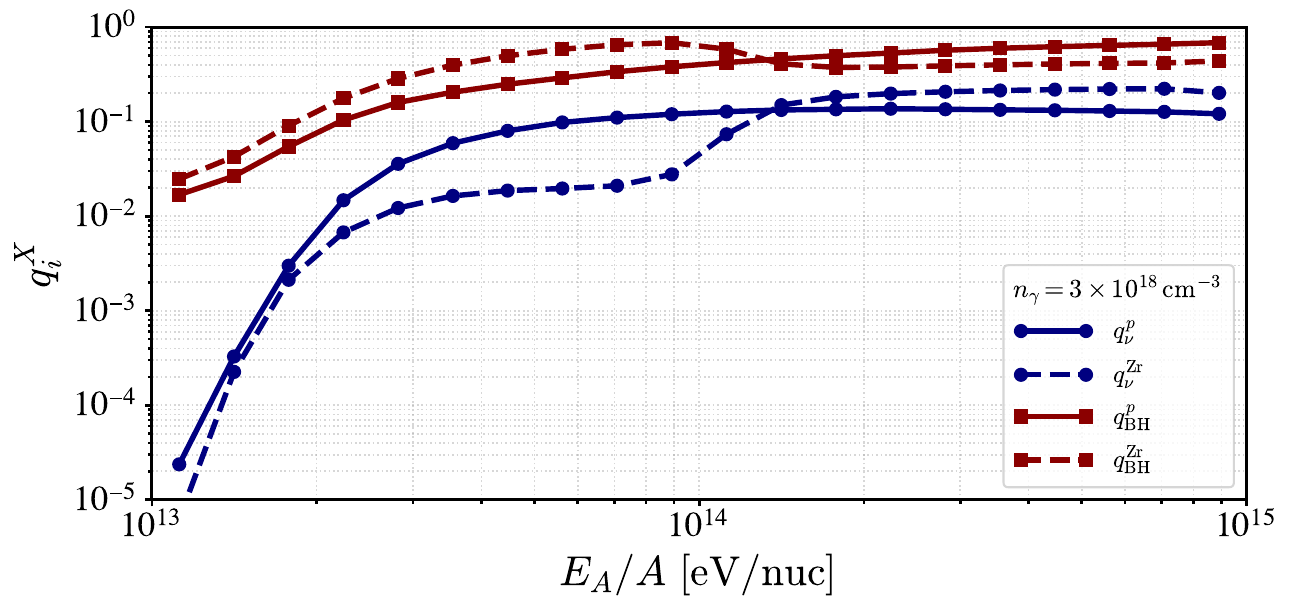}
      \label{fig:Zr_p_binbybin_budgets_doublegauss_ngtot3e18}
    \end{minipage}
  }

  \caption{
  Comparison of the interaction/energy-loss optical depths and the corresponding energy-budget fractions for the double-Gaussian photon field.
  Left column: optical depths for proton and zirconium injection.
  Right column: bin-by-bin neutrino and Bethe--Heitler energy-budget fractions.
  From top to bottom, the total photon densities are
  \(n_\gamma=3\times10^{16}\), \(3\times10^{17}\), and \(3\times10^{18}\,{\rm cm^{-3}}\).
  }
  \label{fig:doublegauss_optdepth_budget_comparison}
\end{figure*}

\begin{figure}[t!]
  \centering
  \includegraphics[width=0.99\columnwidth]{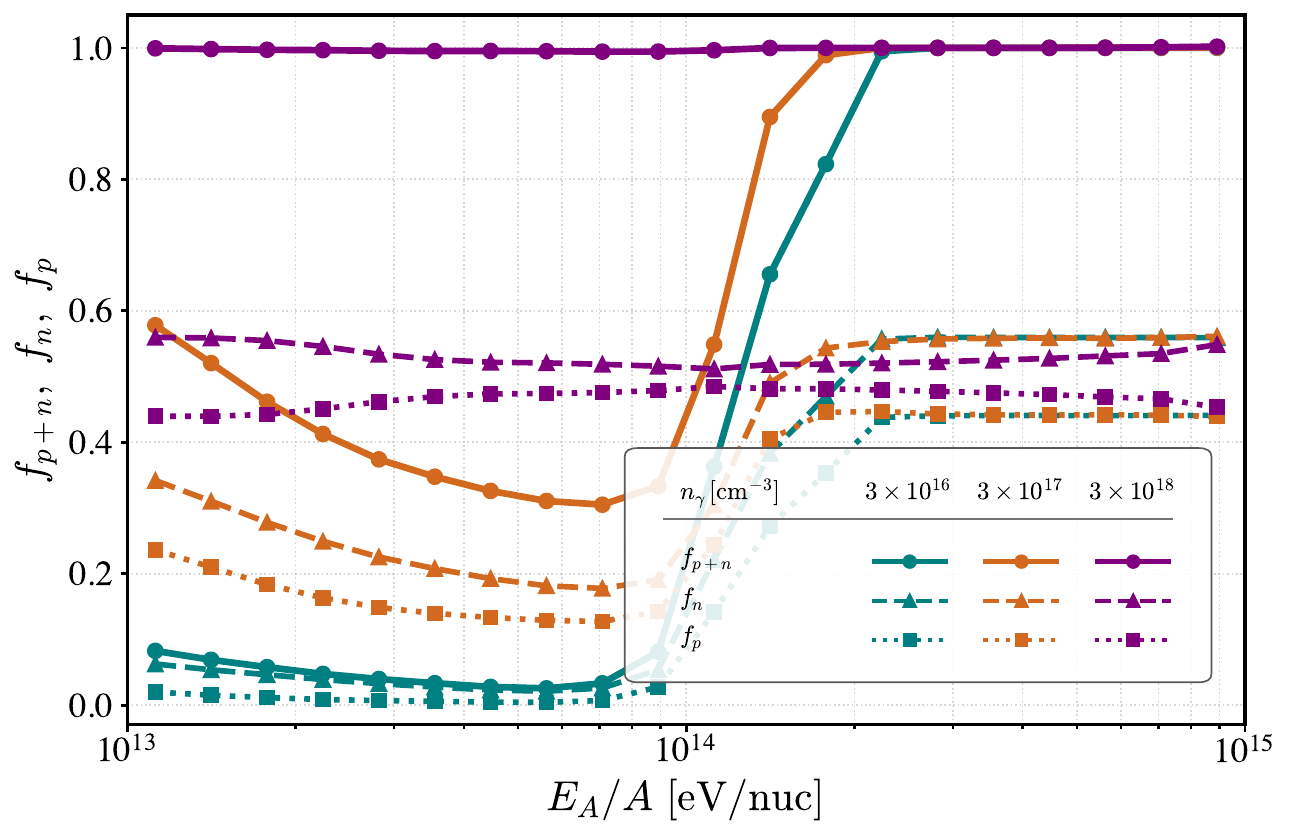}
  \caption{
  Escaped free-nucleon fractions for zirconium injection in the double-Gaussian photon field.
  The curves show \(f_{p+n}\), \(f_n\), and \(f_p\) for the three photon densities considered in Fig.~\ref{fig:doublegauss_optdepth_budget_comparison}.
  }
  \label{fig:Zr_nucleon_fractions_doublegauss}
\end{figure}

\begin{figure*}[t]
\subfloat[]{
\includegraphics[width = 0.32\textwidth]{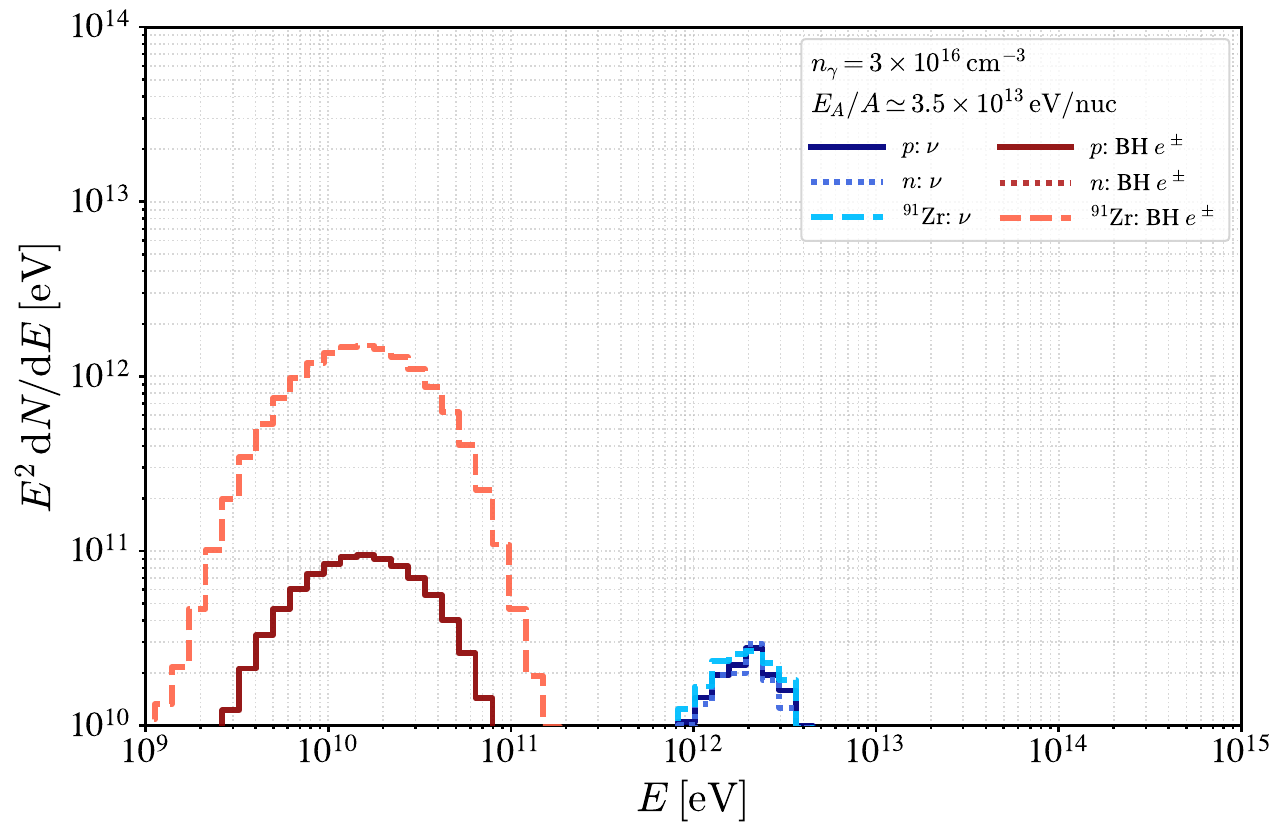}
\label{fig:onebin005_p_n_nu_BH_PM_ngtot3e16}
}
\hfill
\subfloat[]{
\includegraphics[width = 0.32\textwidth]{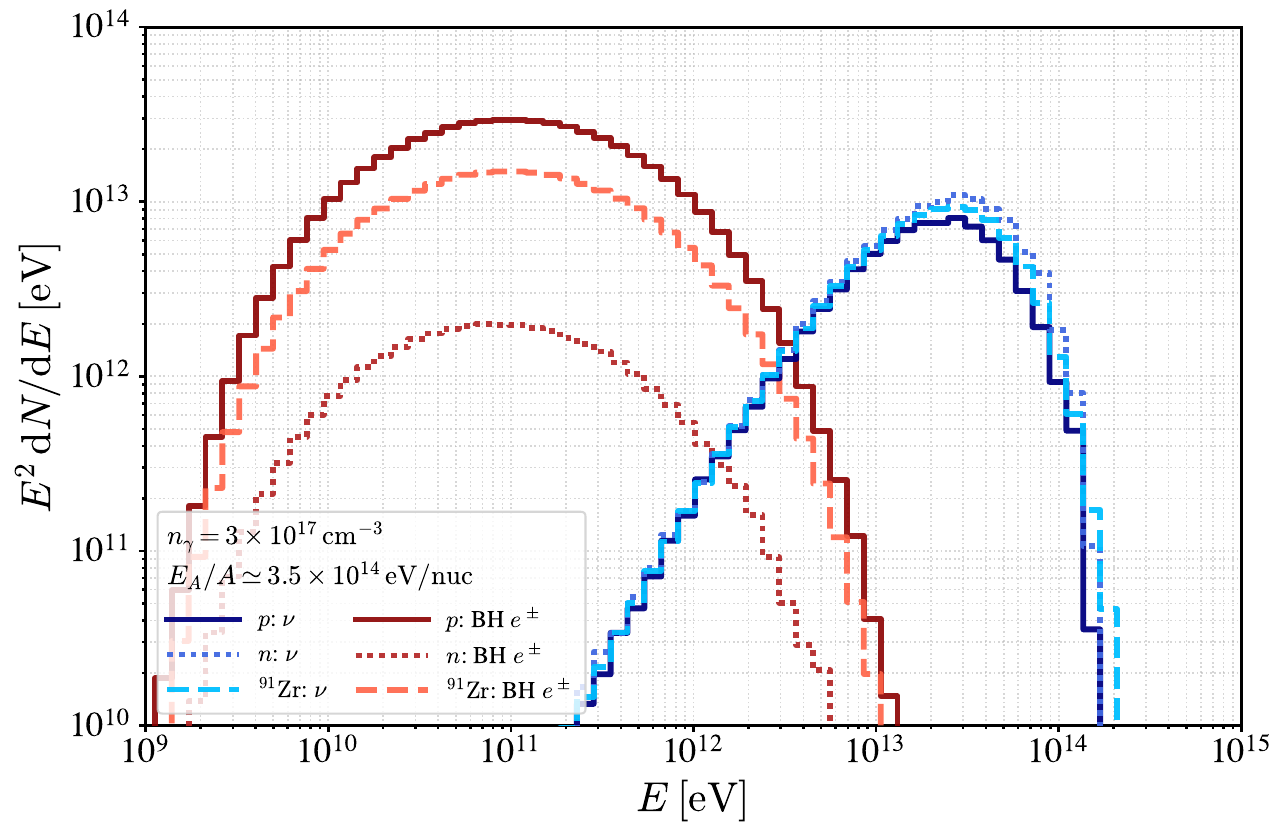}
\label{fig:onebin015_p_n_nu_BH_PM_ngtot3e17}
}
\hfill
\subfloat[]{
\includegraphics[width = 0.32\textwidth]{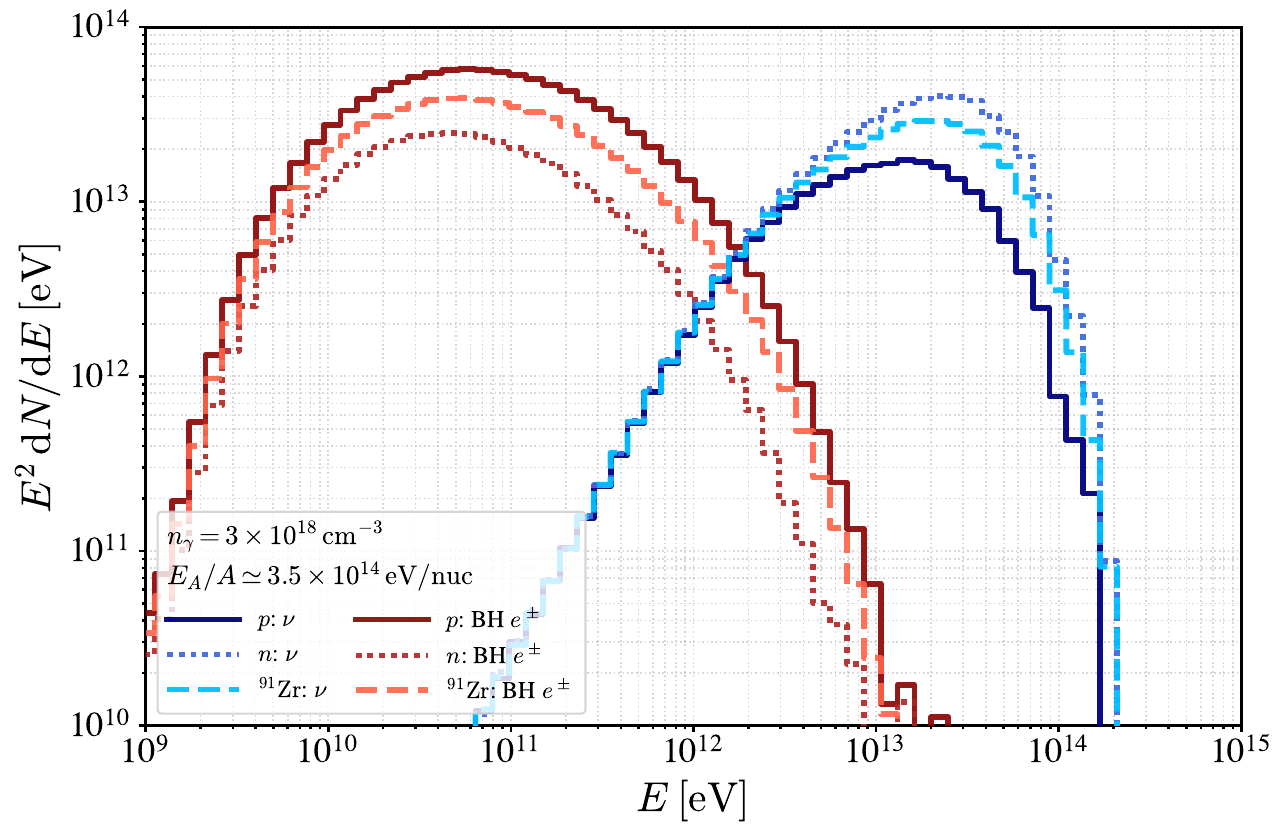}
\label{fig:onebin015_p_n_nu_BH_PM_ngtot3e18}
}
\caption{Photomeson neutrino spectra and Bethe–Heitler electromagnetic secondary spectra from proton, neutron, and zirconium propagation in the double-Gaussian photon field. The spectra are shown at fixed energy per nucleon, $E_A/A\simeq3.5\times10^{13}\,{\rm eV/nuc}$, in the left panel, and $E_A/A\simeq3.5\times10^{14}\,{\rm eV/nuc}$ in the middle and right panels. The target photon densities are $n_\gamma=3\times10^{16}\,{\rm cm}^{-3}$ (left), $3\times10^{17}\,{\rm cm}^{-3}$ (middle), and $3\times10^{18}\,{\rm cm}^{-3}$ (right). For zirconium injection, the spectra are normalized per injected nucleon, allowing for a direct comparison with the proton and neutron cases.}
\label{fig:n_p_single_energy_secondaries}
\end{figure*}

Let us first consider the most rarefied photon setup, $n_{\gamma}=3\times10^{16}~{\rm cm^{-3}}$, whose optical depths are shown in Fig.~\ref{fig:opticaldepths_zr_p_doublegauss_3e16}, focusing on the range $10^{13}\,{\rm eV/nuc}\lesssim E_A/A \lesssim 10^{15}\,{\rm eV/nuc}$. In this regime, the proton BH optical depth remains below unity over the relevant energy range, $\tau_p^{\rm BH}<1$. Only for sufficiently heavy nuclei the nuclear BH optical depth approaches unity, as illustrated for Zr in the figure. Thus, both protons and heavy nuclei can traverse the source without losing a sizable fraction of their energy via BH pair production. As a consequence, when PM production on the energetic ``corona'' background becomes operative, the neutrino energy fraction is approximately universal when compared at fixed energy per nucleon, i.e. $q^p_\nu(E)\simeq q^{\rm Zr}_\nu(E_A/A)$, as shown in Fig.~\ref{fig:Zr_p_binbybin_budgets_doublegauss_ngtot3e16}.
However, the composition of the escaping nuclear component is non-trivial. At lower energies, photodisintegration proceeds mainly through the less abundant coronal photons ($\tau_A^{\rm PD}\sim 10$) and is less complete before escape. A non-negligible fraction of the injected energy therefore remains in bound nuclear fragments, which have lower optical depth to PD than their parent nucleus. These fragments still experience the enhanced nuclear BH loss, so the BH energy fraction of Zr increases relative to the proton case, as seen in the low-energy part of Fig.~\ref{fig:Zr_p_binbybin_budgets_doublegauss_ngtot3e16}. At $E_A/A\gtrsim10^{14}\,{\rm eV/nuc}$, Zr interacts efficiently with the abundant UV component through photodisintegration process, reaching $\tau_{\rm Zr}^{\rm PD}\gtrsim10^2$. The nucleus is then rapidly converted into free nucleons ($f_{p+n}\approx 1$), as also indicated by the escaped nucleon fractions in Fig.~\ref{fig:Zr_nucleon_fractions_doublegauss}. Since the PM of nucleons is inefficient ($\tau_p^{\rm PM}\sim 0.05 - 0.1$) most of the neutrons released by the photodisintegration escape the source without further interaction while only a small fraction undergoes at most a single photomeson\footnote{For a Poisson process with optical depth $\tau$, the probability of exactly one interaction is $P_1=\tau e^{-\tau}$. For $\tau=0.05-0.1$, this gives $P_1\simeq 5-10\,\%$, while the probability of two or more interactions is negligible.}. Since a significant part of this nucleon population is neutral and neutrons do not undergo BH losses, the BH energy fraction for Zr is smaller than that for proton injection, $q_{\rm BH}^{\rm Zr}\lesssim q_{\rm BH}^{p}$, although both remain modest and below $10\%$. The relevant yields associated to this rarefied regime are shown in Fig.~\ref{fig:onebin005_p_n_nu_BH_PM_ngtot3e16}, for $E_A/A\simeq3.5\times10^{13}\,{\rm eV/nuc}$: the neutrino spectra (per nucleon) are very close for a proton, neutron or Zr injection, as manifestation of the quasi-universality we have mentioned. The BH yields for the case of nuclei are stronger than the proton case, signaling a larger e.m. to neutrino flux ratio for nuclei at low energy. 
Note the strongly suppressed BH losses of neutron injection, since this requires the relatively rare PM event accompanied by a $n\to p$ conversion, in turn followed by losses in the remaining path.

%%%%%%%%%%%%%%%%%%%%%%%%%%%%%%%
%%%%%%%%%%%% FIG. 2 %%%%%%%%%%%
%%%%%%%%%%%%%%%%%%%%%%%%%%%%%%%

%%%%%%%%%%%%%%%%%%%%%%%%%%%%%%%
%%%%%%%%%%%% FIG. 3 %%%%%%%%%%%
%%%%%%%%%%%%%%%%%%%%%%%%%%%%%%%

We next consider the intermediate photon density benchmark, $n_{\gamma}=3\times10^{17}\,{\rm cm^{-3}}$, whose optical depths are shown in Fig.~\ref{fig:opticaldepths_zr_p_doublegauss_3e17}. In this case, the proton BH optical depth reaches values of order unity at its peak value, while the BH optical depth of heavy nuclei is already sizable over most of the range, apart from the lowest energies. At lower energies, the qualitative behaviour resembles that in the rarefied photon background. Photodisintegration proceeds mainly through the less abundant coronal component and is not complete before escape. A significant fraction of the injected energy therefore remains in bound nuclear fragments, as shown by the escaped nucleon fractions in Fig.~\ref{fig:Zr_nucleon_fractions_doublegauss}. The important difference with respect to the rarefied case is that the BH losses of these fragments are now sizable compared to those of protons. As a result, a larger fraction of the injected nuclear energy is channeled into BH pairs rather than neutrinos, leading to $q_\nu^{\rm Zr}<q_\nu^p$ and $q_{\rm BH}^{\rm Zr}>q_{\rm BH}^p$, as can be seen in Fig.~\ref{fig:Zr_p_binbybin_budgets_doublegauss_ngtot3e17}. Toward the lowest energies shown, close to the PM threshold on the coronal photon field, BH losses again become negligible and the approximate universality of the neutrino yield is recovered.
This behaviour changes at higher energies, where photodisintegration is very fast. Since the nucleon PM optical depth is only moderate, $\tau_{p,n}^{\rm PM}\sim1$, the free neutrons released by photodisintegration can contribute to neutrino production through an $\mathcal{O}$($\sim$ few) number of PM interactions, while avoiding continuous BH losses before their conversion into charged baryons. Moreover, neither the protons directly released by photodisintegration nor those produced after PM interactions of neutrons lose a large fraction of their energy through BH pair production ($\tau_p^{\rm BH}\sim1$). This leads to a mild enhancement of the neutrino energy fraction in the Zr-injection case, $q_\nu^{\rm Zr}\gtrsim q_\nu^p$, compared to the lower density benchmark.
The BH energy fraction follows the complementary trend. In proton injection, the increase of $\tau_p^{\rm BH}$ toward unity enhances the energy transfer to pairs. In Zr injection, however, an increasing fraction of the energy is carried by neutrons after photodisintegration, reducing the fraction of the cascade energy exposed to BH losses. Once the cascade reaches the free-nucleon stage, this neutral component partially compensates for the enhanced BH losses experienced by charged nuclear fragments at earlier stages of propagation. As a result, $q_{\rm BH}^{\rm Zr}/q_{\rm BH}^p$ decreases at high energy, and its value is smaller than in the more rarefied benchmark discussed above.
Characteristic yields associated with this intermediate regime are shown in Fig.~\ref{fig:onebin015_p_n_nu_BH_PM_ngtot3e17}, for $E_A/A\simeq3.5\times10^{14}\,{\rm eV/nuc}$. The BH spectrum for proton injection now exceeds the Zr one, reflecting the fact that, once photodisintegration has efficiently released free nucleons, only the charged component remains exposed to continuous BH cooling. At the same time, a non-negligible BH yield also appears for neutron injection, due to neutron PM interactions producing charged final-state baryons which subsequently undergo BH losses. The neutrino spectra, normalized per nucleon, show the complementary trend: the proton-injection case is slightly lower than the neutron one, because BH cooling has started to reduce the energy available for PM production in charged projectiles.

The final benchmark corresponds to the densest photon field considered here, $n_{\gamma}=3\times10^{18}\,{\rm cm^{-3}}$, whose optical depths are shown in Fig.~\ref{fig:opticaldepths_zr_p_doublegauss_3e18}. In this case, BH losses are efficient already for protons, $\tau_p^{\rm BH}>1$, and therefore even more so for heavy nuclei, $\tau_A^{\rm BH}>\tau_p^{\rm BH}>1$. The system is thus in a regime where electromagnetic energy losses are no longer a perturbation, even for a pure-proton injection.
At lower energies, photodisintegration is still sizable, but it proceeds mainly through the less abundant coronal photon field. As a result, the transition to the free-nucleon regime requires a longer propagation path, and the injected nucleus spends a non-negligible fraction of its history in the form of heavy nuclear fragments. During this stage, the enhanced BH losses of these fragments reduce the energy available for neutrino production. Consequently, the neutrino energy fraction in the Zr-injection case is smaller than for proton injection, $q_\nu^{\rm Zr}<q_\nu^p$, as shown in Fig.~\ref{fig:Zr_p_binbybin_budgets_doublegauss_ngtot3e18}. At the same time, $q_{\rm BH}^{\rm Zr}$ remains larger than $q_{\rm BH}^{p}$ in this low-energy part, reflecting the stronger BH cooling of charged nuclear fragments. However, because protons themselves now suffer substantial BH losses, the separation between the two BH energy fractions is less pronounced than in the intermediate density benchmark.
At higher energies, photodisintegration becomes even more rapid, and the cascade reaches the free-nucleon regime earlier in the source. The system therefore spends a larger fraction of its propagation in the nucleon-dominated stage. In this dense environment, the PM optical depth of nucleons is also sizable, $\tau_{p,n}^{\rm PM}>1$. The subsequent evolution is therefore not described by free neutron escape, but rather by repeated proton–neutron conversions through PM interactions. Schematically, a pure-proton injection can undergo charge-exchange sequences of the form
\begin{equation}
    p\to n\to p\to n\to\cdots ,
\end{equation}
whereas a fully disintegrated nuclear injection starts from a mixed free-nucleon population,
\begin{equation}
    (p,n)\to(n,p)\to(p,n)\to(n,p)\to\cdots .
\end{equation}
In the limit of large PM optical depth, this sequence tends to make the charged and neutral nucleon populations more symmetric, thereby reducing the distinction between proton and nuclear injection.

There remains, however, an important asymmetry at the beginning of the sequence. In the nuclear-injection case, a sizable fraction of the energy is released directly into neutrons by photodisintegration. This neutral component is not exposed to BH losses before its first PM interaction. Since PM  production itself transfers a sizable fraction of the projectile energy to secondary particles, the energy carried by the leading baryon after this first interaction is already reduced. By contrast, in the pure-proton case, the injected energy is carried by charged particles from the beginning and is therefore subject to BH losses already on the first leg of the sequence. Thus, even when PM interactions are efficient and proton–neutron cycling develops, neutron-initiated sequences remain less affected by BH losses than proton-initiated ones. In this regime, nuclear injection is therefore more efficient than proton injection in neutrino rather than BH pair production, provided that the comparison is made at a fixed energy per nucleon. Correspondingly, at high energies one finds $q_\nu^{\rm Zr}>q_\nu^p$, while $q_{\rm BH}^{\rm Zr}$ is reduced relative to $q_{\rm BH}^{p}$, as shown in Fig.~\ref{fig:Zr_p_binbybin_budgets_doublegauss_ngtot3e18}.
Characteristic yields associated with this dense-target regime are shown in Fig.~\ref{fig:onebin015_p_n_nu_BH_PM_ngtot3e18}, for $E_A/A\simeq3.5\times10^{14}\,{\rm eV/nuc}$. The BH spectra are now closer to each other than in the intermediate-density case, reflecting the combined effect of efficient BH cooling and repeated PM-induced neutron–proton conversion. The distinction between initially charged and neutral projectiles is therefore partially washed out in the electromagnetic channel. The neutrino spectra, normalized per nucleon, retain a clearer hierarchy, however, because the ``first interaction'' effect previously described.

%%%%%%%%%%%%%%%%%%%%%%%%%%%%%%%%%%%%%%%%
%%%%%%%%%%%%%%%  FIG 5  %%%%%%%%%%%%%%%%

%%%%%%%%%%%%%%%%%%%%%%%%%%%%%%%%%%%%%%
%%%%%%%%%%%%%%%%%%%%%%%%%%%%%%%%%%%%%%

A remark is in order. In all the discussions, we assume fixed energy per nucleon, so that a proton of energy $E_p$ is compared with a nucleus of total energy $E_A=A E_p$. This choice is appropriate for comparing neutrino production at the same characteristic neutrino energy, since the relevant interaction kinematics are controlled mainly by the nucleon Lorentz factor, $\Gamma_A\simeq E_A/(A m_p)$, where $m_p$ is the proton mass. From the acceleration point of view, however, the maximum energy of a charged particle in the same acceleration environment is expected to scale with rigidity, and hence with $Z$, rather than with $A$. Therefore, although heavy nuclei can be more efficient neutrino producers at fixed $E_A/A$ in the dense and nucleon-dominated regime, this does not imply that they are generically more efficient than protons once acceleration characteristics are considered.

The simplified disk–corona setup discussed in this section isolates the basic physical mechanisms that control the relative neutrino and electromagnetic yields of nuclear and proton injection. The key ingredients are the degree of photodisintegration before escape, the amount of energy lost by charged fragments through BH pair production, and the role of the neutral nucleon component once the cascade reaches the free-nucleon regime. In the next section, we apply this physical picture to a more realistic disk–corona model of NGC~1068, based on \cite{Das:2024vug}. We specify the corresponding photon field and injection spectra, and then use the Monte Carlo output to quantify the escaped composition and the resulting multimessenger emissions.

%%%%%%%%%%%%%%%%%%%%%%%%%%%%%%%
%%%%%%%%%%%%%%%%%%%%%%%%%%%%%%%
%%%%%%%%%%%%%%%%%%%%%%%%%%%%%%%
\section{The NGC~1068 model \& results}\label{sec:model}
%%%%%%%%%%%%%%%%%%%%%%%%%%%%%%%
%%%%%%%%%%%%%%%%%%%%%%%%%%%%%%%

For the NGC~1068 application, we adopt the disk–corona radiation field of ref.~\cite{Das:2024vug}. The central black-hole mass is fixed to $M=10^7\,M_\odot$, and the particle production region is modeled as a homogeneous sphere of radius $R$. We keep the spectral shape and luminosity of the disk and corona components fixed to the reference model, so that changing $R$ also rescales the photon density as $R^{-2}$. Therefore, the source size is the parameter that controls the effective photonuclear and electromagnetic column depths in the NGC~1068 setup.

\begin{figure*}[t]
\subfloat[]{
\includegraphics[width = 0.48\textwidth]{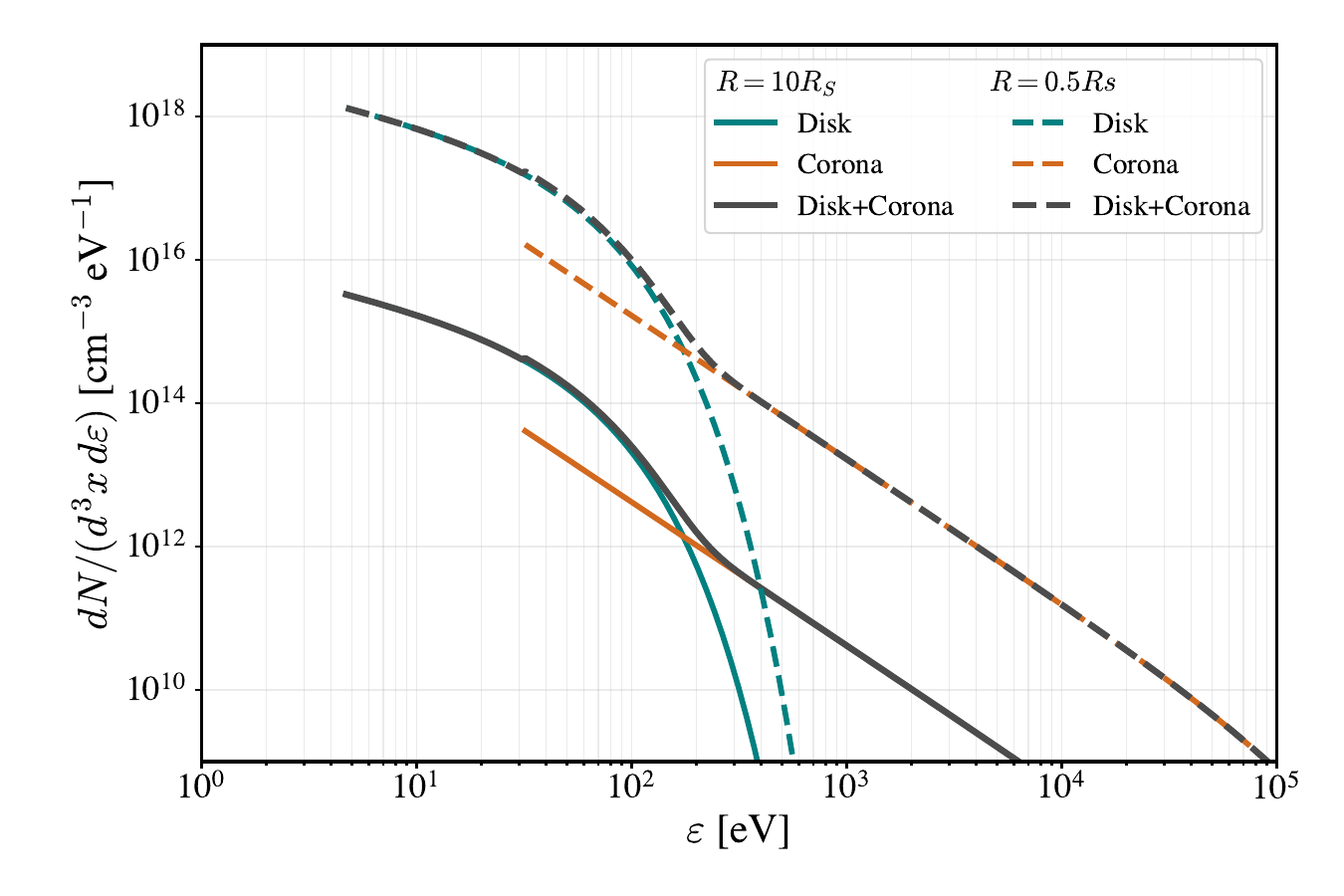}
\label{fig:corona_disk_photontarget}
}
\hfill
\subfloat[]{
\includegraphics[width = 0.48\textwidth]{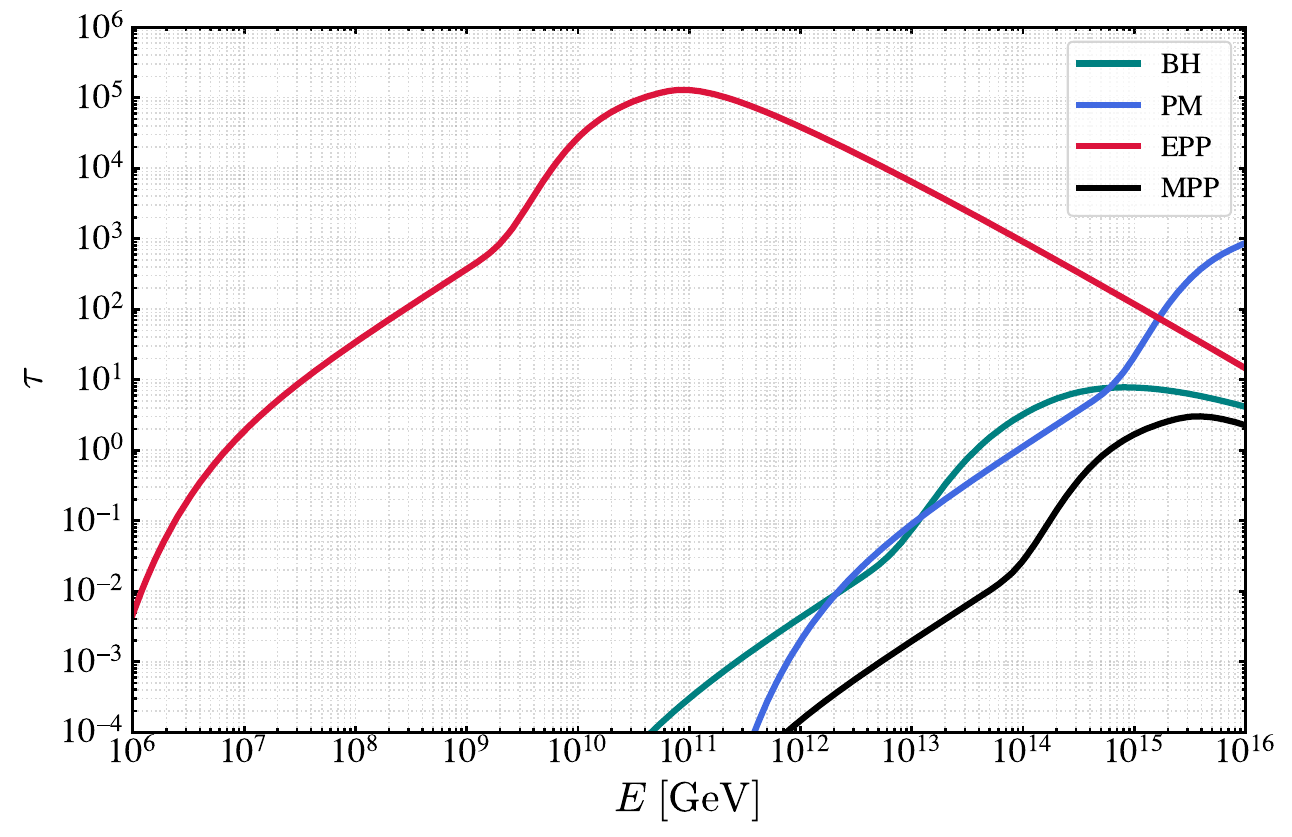}
\label{fig:tau_2405_09332}
}
\caption{(a) Target photon field of the corona+disk model based on Eqs. 2 and 3 of ref.~\cite{Das:2024vug} for $R=10R_S$ and $R=0.5R_S$ source sizes, depicted by solid and dashed curves respectively. (b) Proton optical depths for $R=10R_S$, using the same photon field and escape prescription of ref.~\cite{Das:2024vug}, reproducing their Fig.~1.}
\label{fig:NGC1068_model}
\end{figure*}

Fig.~\ref{fig:NGC1068_model} summarizes the target field used in the simulations. In Fig.~\ref{fig:corona_disk_photontarget} we show the disk and coronal photon spectra for two representative source sizes, $R=10R_S$ (solid lines) and $R=0.5R_S$ (dashed lines), illustrating the denser photon field for more compact sources. For reference, the $10R_S$ source size leads to the characteristic photon number density $n_\gamma\simeq4\times10^{16}\,\mathrm{cm^{-3}}$, while for $0.5R_S$ source we find $n_\gamma\simeq1.6\times10^{19}\,\mathrm{cm^{-3}}$. In both cases the target fields are dominated by photons with energies of a few eV. Fig.~\ref{fig:tau_2405_09332} provides a consistency check by reproducing the proton mean free paths shown in Fig.~1 of ref.~\cite{Das:2024vug} for $R=10R_S$, using the same photon field and escape prescription.

%%%%%%%%%%%%%%%%%%%%%%%%%%%%%%%%%%%%%%%%
%%%%%%%%%%%%%%%  FIG 6  %%%%%%%%%%%%%%%%

%%%%%%%%%%%%%%%%%%%%%%%%%%%%%%%%%%%%%%
%%%%%%%%%%%%%%%%%%%%%%%%%%%%%%%%%%%%%%

The source sizes explored in this work are chosen to lie within the range favored by multimessenger and multiwavelength constraints of ref.~\cite{Das:2024vug}. In particular, electromagnetic cascade constraints, together with the recent Fermi-LAT analysis of the gamma-ray emission from NGC~1068~\cite{Ajello:2023hkh}, imply $R\lesssim 15\,R_S$ for strongly magnetized (so-called low-$\beta$) plasma, with even tighter bound $R\lesssim 3\,R_S$ in weakly magnetized (high-$\beta$) environments~\cite{Das:2024vug}. In the following, we focus on $0.5\,R_S<R<10\,R_S$, using this interval as the compact disk--corona benchmark range for NGC~1068.

We assume the injected CR spectrum to be
\begin{equation}\label{eq:inj_spec}
    \frac{\mathrm{d}\dot{N}_A}{\mathrm{d}E_A}
    =
    \mathcal{N}_A
    E_A^{-\gamma}
    \exp\!\left(-\frac{E_A}{E_{A}^{\rm cut}}\right),
    \qquad
    E_A\geq E_{A}^{\rm min},
\end{equation}
where the spectral index is fixed to $\gamma=2$. For protons we adopt the benchmark values $E_p^{\rm min}=10\,{\rm TeV}$ and $E_p^{\rm cut}=30\,{\rm TeV}$, as in ref.~\cite{Das:2024vug}. In the disk--corona photon field, this range yields neutrinos in the TeV-band, in accordance with the IceCube signal from NGC~1068. For nuclear injection, we use the fixed-$E_A/A$ comparison introduced in Sec.~\ref{sec:multimessenger} and scale the injection interval with the mass number,
\begin{equation}
    E_A^{\rm min}=A E_p^{\rm min},\qquad
E_A^{\rm cut}=A E_p^{\rm cut}.
\end{equation}
This ensures that different injected compositions populate the same characteristic neutrino energy range.

The normalization $\mathcal{N}_A$ is fixed by the injected cosmic-ray luminosity $L_{\rm CR}$, 
\begin{equation}
    L_{\rm CR}
    =
    \int_{E_{\rm min}^{A}}^{\infty}
    E_A\,
    \frac{\mathrm{d}\dot{N}_A}{\mathrm{d}E_A}~
    \mathrm{d}E_A~.
\end{equation}
For each source size and injected composition, $L_{\rm CR}$ is anchored to the best-fit of NGC~1068 flux measured by IceCube. The accompanying electromagnetic secondaries are propagated through the same disk–corona photon field using the cascade treatment described in Sec.~\ref{sec:nuc_cascades}. The injected electromagnetic energy is thereby reprocessed toward lower photon energies, yielding a MeV-GeV gamma-ray flux that can be tested against the available measurements and upper limits.

Before discussing the full multimessenger output, we first quantify how efficiently the injected nuclei are processed in the NGC~1068 radiation field. This step is useful because, as discussed in Sec.~\ref{sec:multimessenger}, the degree of photodisintegration before escape determines the efficiency of the relevant energy loss processes and hence the partition of the injected energy to neutrinos and electromagnetic secondaries. Without repeating the arguments presented in Sec.\ref{sec:multimessenger}, in this section we show the results for the adopted disk–corona model of NGC~1068.

%%%%%%%%%%%%%%%%%%%%%%%%%%%%%%%
%%%%%%%%%%% FIG. 7 %%%%%%%%%%%%
\begin{figure}[t]
  \centering
  \includegraphics[width=0.99\columnwidth]{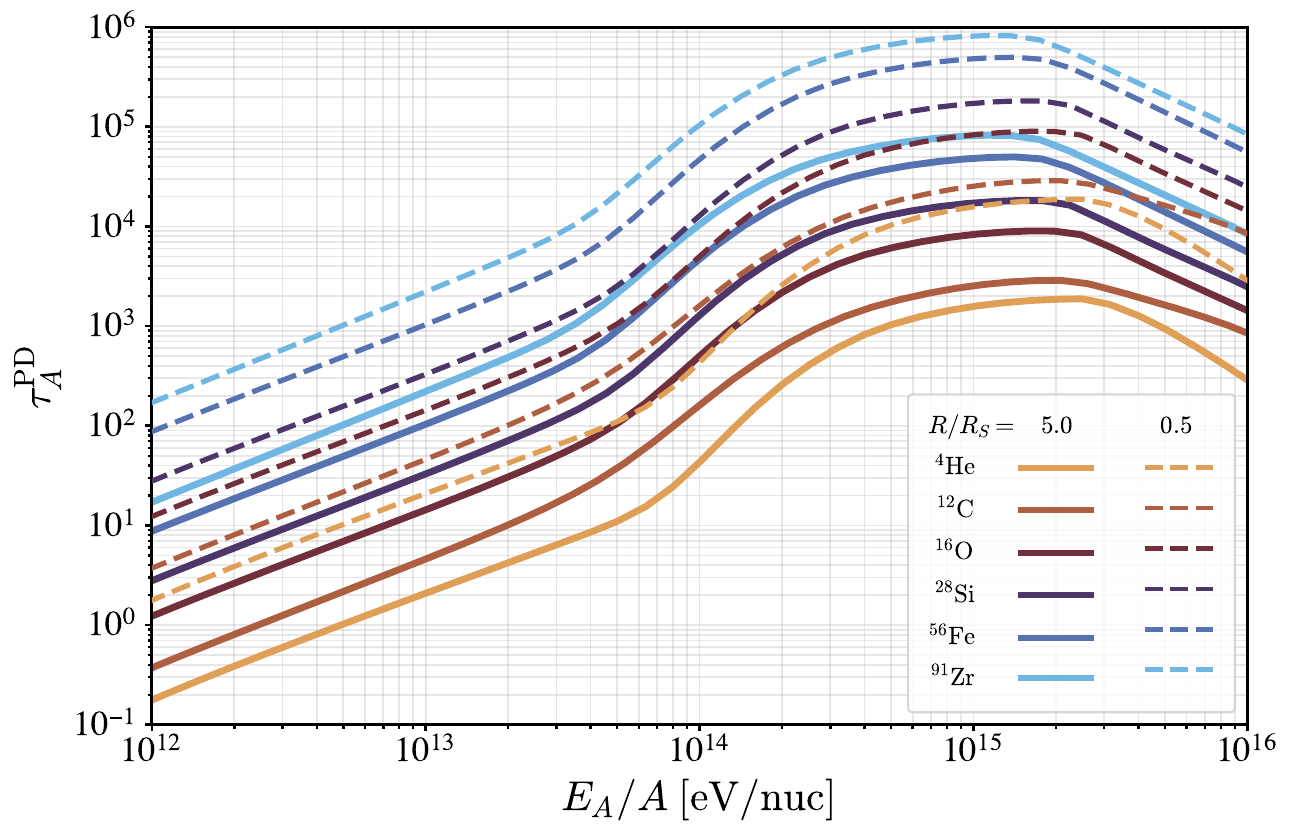}
  \caption{Photodisintegration optical depth, $\tau^{\rm PD}_A=R/\lambda^{\rm PD}_{A}$, for representative nuclei propagating in the disk--corona radiation field of NGC~1068 based on ref.~\cite{Das:2024vug}. The curves are shown as a function of energy per nucleon, $E_A/A$  for production sizes of $R=5R_S$ (solid lines) and $R=0.5R_S$ (dashed lines), and for different nuclear species (color-coded).
}
  \label{fig:pd_optical_depth}
\end{figure}
%%%%%%%%%%%%%%%%%%%%%%%%%%%%%%%
%%%%%%%%%%%%%%%%%%%%%%%%%%%%%%%

Fig.~\ref{fig:pd_optical_depth} shows the photodisintegration optical depths for several nuclei (color-coded similar to Fig.~\ref{fig:pd_optical_depth_doublegauss}) as function of $E_A/A$, for a relatively rarefied source of size $R=5R_S$ (solid lines) and a compact and dense source with $R=0.5R_S$ (dashed lines). In the injection range relevant for the IceCube-normalized spectra, $E_A/A\simeq10\,\text{--}\,30\,{\rm TeV}$, carbon has only a moderate photodisintegration optical depth for $R=5R_S$, $\tau^{\rm PD}_{\rm C}\sim{\cal O}({\rm few})$, whereas zirconium is already very optically thick, with $\tau^{\rm PD}_{\rm Zr}\sim{\cal O}(10^2\text{--}10^3)$. Reducing the source size to $R=0.5R_S$ increases the photon density and correspondingly raises the photodisintegration optical depths for all injected species.

%%%%%%%%%%%%%%%%%%%%%%%%%%%%%%%
%%%%%%%%%%%% FIG. 8 %%%%%%%%%%%
\begin{figure}[t]
  \centering
  \includegraphics[width=0.99\columnwidth]{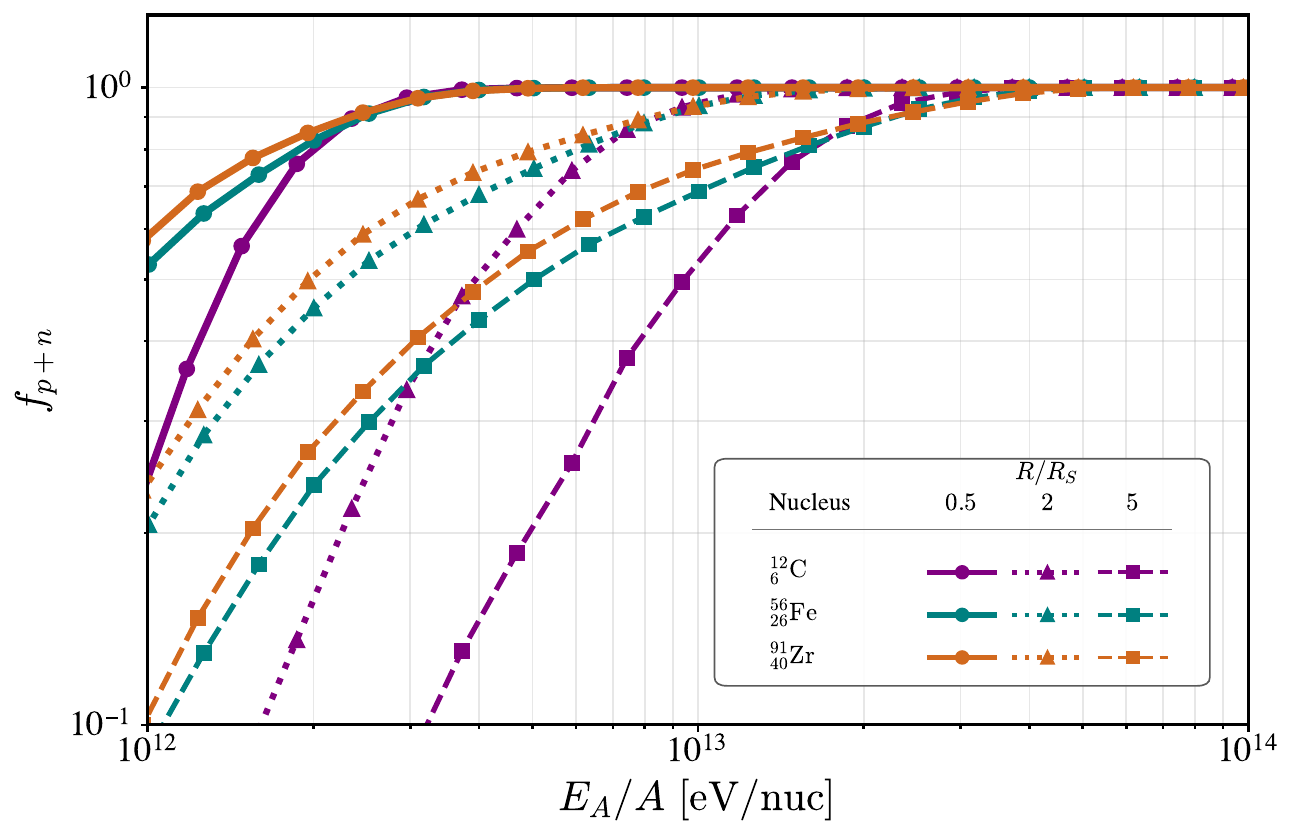}
  \caption{Escaped free-nucleon fraction, $f_{p+n}$, as a function of energy per nucleon, $E_A/A$, for pure carbon, iron, and zirconium injection in the NGC1068 disk–corona radiation field based on ref.~\cite{Das:2024vug}. The solid, dashed, and dotted curves correspond to source sizes $R/R_S=0.5$, $5.0$, and $2.0$, respectively.}
  \label{fig:nucleonfrac_vs_energy}
\end{figure}
%%%%%%%%%%%%%%%%%%%%%%%%%%%%%%%
%%%%%%%%%%%%%%%%%%%%%%%%%%%%%%%

This opacity trend is directly reflected in the escaping composition. Fig.~\ref{fig:nucleonfrac_vs_energy} shows the escaped free-nucleon fraction, $f_{p+n}$, as a function of $E_A/A$ for pure carbon, iron, and zirconium injection, and for different source sizes. The transition to $f_{p+n}\simeq1$ occurs at lower energies for smaller $R$, as expected from the corresponding increase in the photon density. At fixed $R$, heavier nuclei, such as zirconium, reach the nucleon-dominated limit more readily than lighter species, reflecting their faster photodisintegration.

%%%%%%%%%%%%%%%%%%%%%%%%%%%%%%%
%%%%%%%%%%%% FIG. 9 %%%%%%%%%%%
\begin{figure}[t]
  \centering
  \includegraphics[width=0.99\columnwidth]{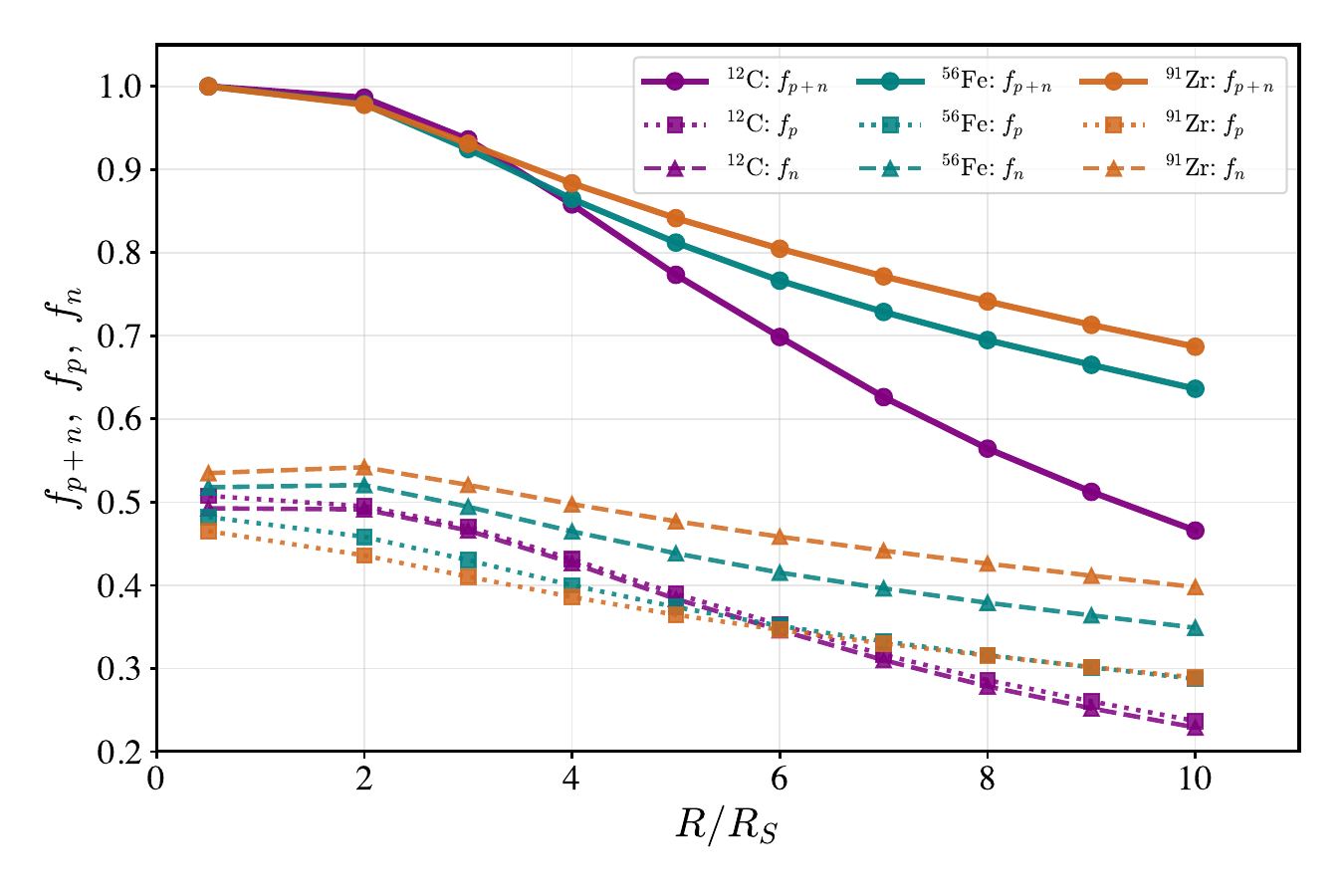}

  \caption{Escaped free-nucleon fractions ($f_p$, $f_n$, and $f_{p+n}$), for carbon, iron, and tungsten injection in the benchmark NGC~1068 models, shown as a function of source size over the range $0.5<R/R_S<10$. The quantity $f_{p+n}$ is computed after integrating over the adopted injection spectrum for each nuclear species. See the text for the details of the heavy-nucleus injection spectra.}
  \label{fig:nucfrac_vs_r}
\end{figure}
%%%%%%%%%%%%%%%%%%%%%%%%%%%%%%%
%%%%%%%%%%%%%%%%%%%%%%%%%%%%%%%

In a complementary way, Fig.~\ref{fig:nucfrac_vs_r} shows the escaped free-nucleon fraction after propagation in the full NGC~1068 setup, as a function of the source size $R$. Unlike Fig.~\ref{fig:nucleonfrac_vs_energy}, which displays the behavior at fixed energy per nucleon, this figure summarizes the net outcome for the injected spectrum of Eq.~\eqref{eq:inj_spec}. It therefore provides a direct measure of whether the escaping baryonic component is nucleon-dominated or still contains a substantial bound-fragment contribution. 

For sufficiently compact regions, $R\lesssim 2R_S$, all injected compositions approach $f_{p+n}\simeq1$, indicating that the escaping baryonic component is nearly fully nucleon dominated. As $R$ increases, the photonuclear column depth decreases and the nuclear cascade is interrupted earlier, so that a larger fraction of the escaping mass remains bounded in nuclear fragments. The transition is composition dependent: heavier nuclei remain closer to the nucleon-dominated limit over a wider range of radii, consistent with their larger PD depths. Fig.~\ref{fig:nucfrac_vs_r} therefore provides a useful diagnostic of which propagation regime is realized for each source size and injected composition.

\begin{figure*}[t]
\centering
\subfloat[]{\includegraphics[width=0.33\textwidth]{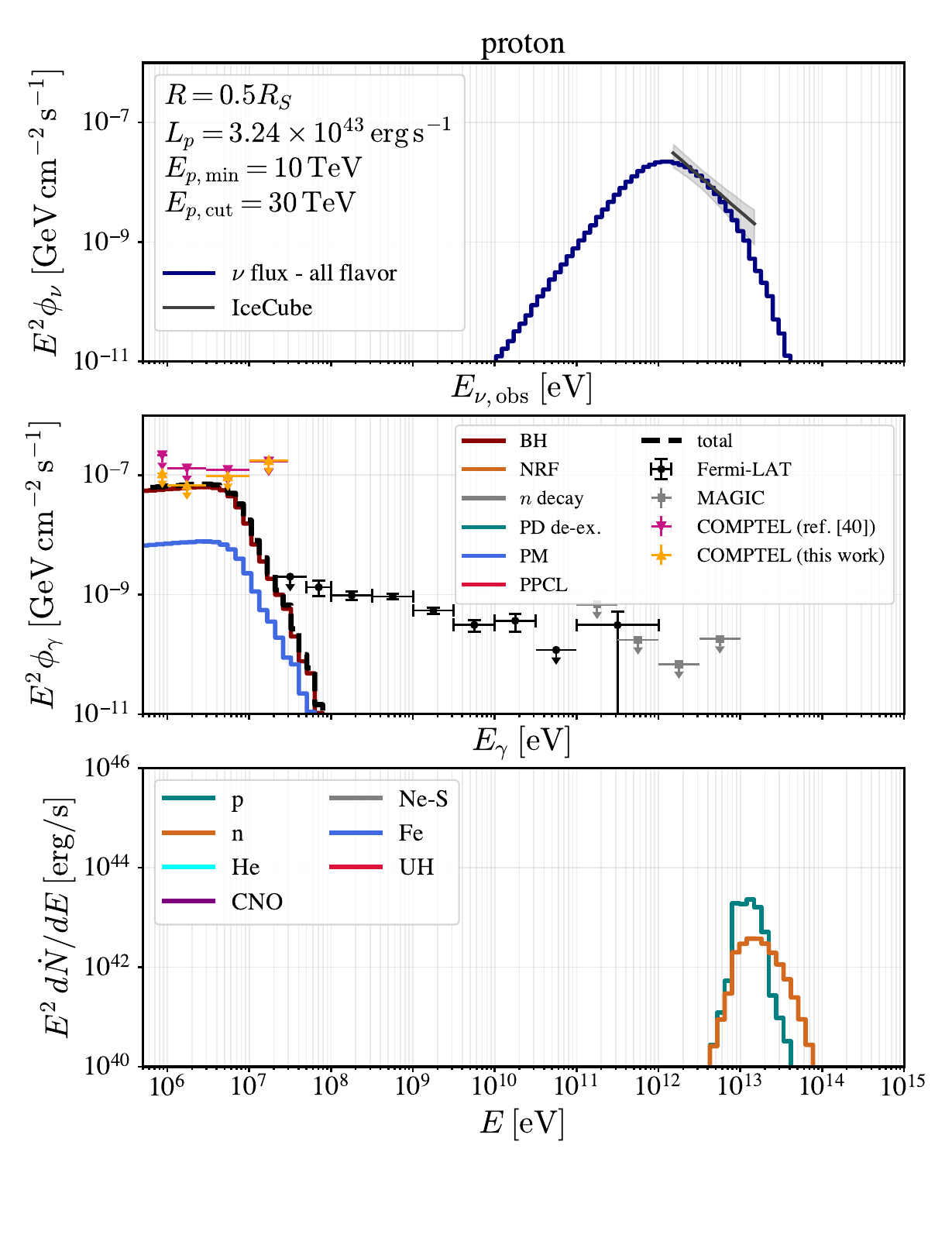}\label{fig:multimessenger_proton_R0.5}}
\hfill
\subfloat[]{\includegraphics[width=0.33\textwidth]{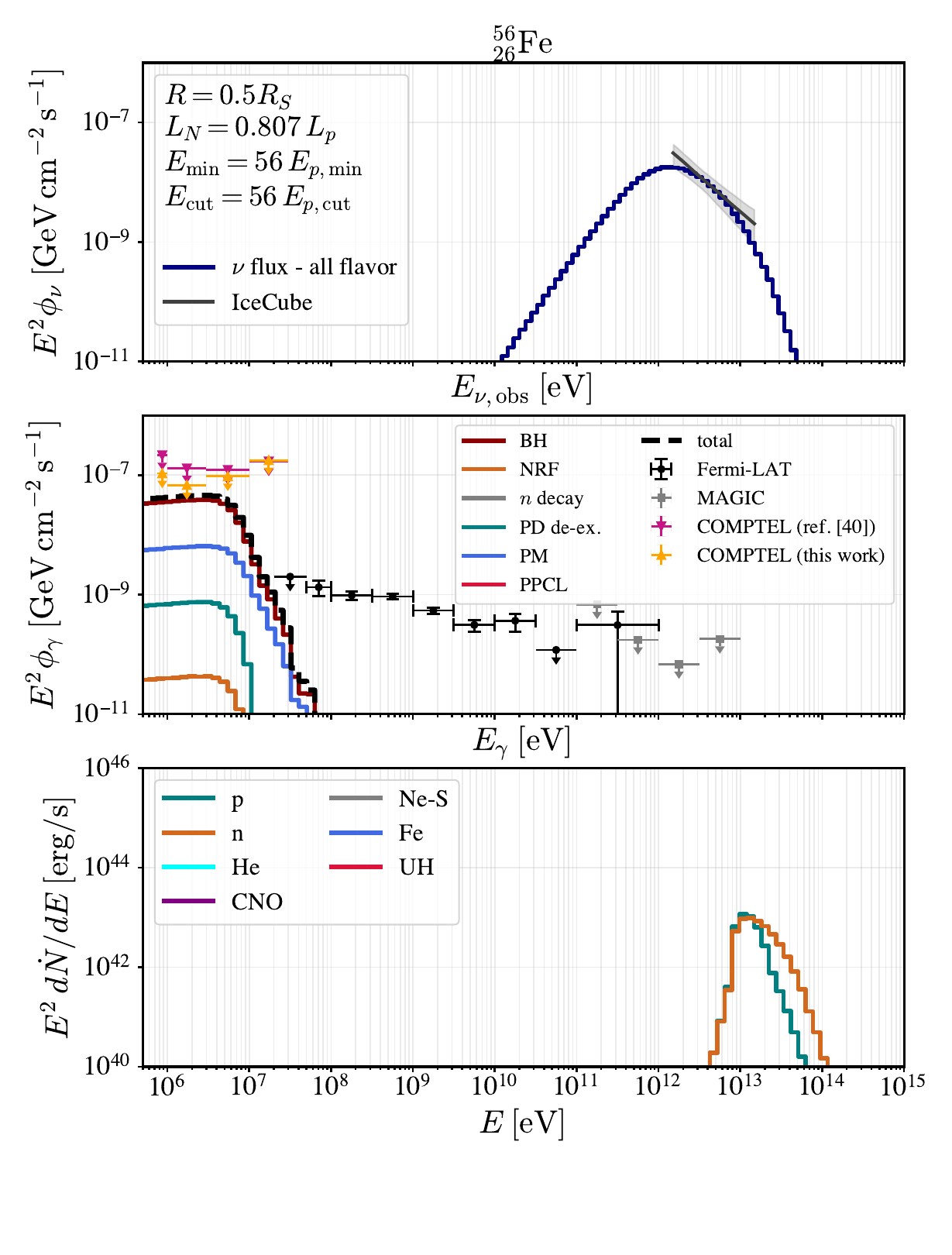}\label{fig:multimessenger_iron_R0.5}}
\hfill
\subfloat[]{\includegraphics[width=0.33\textwidth]{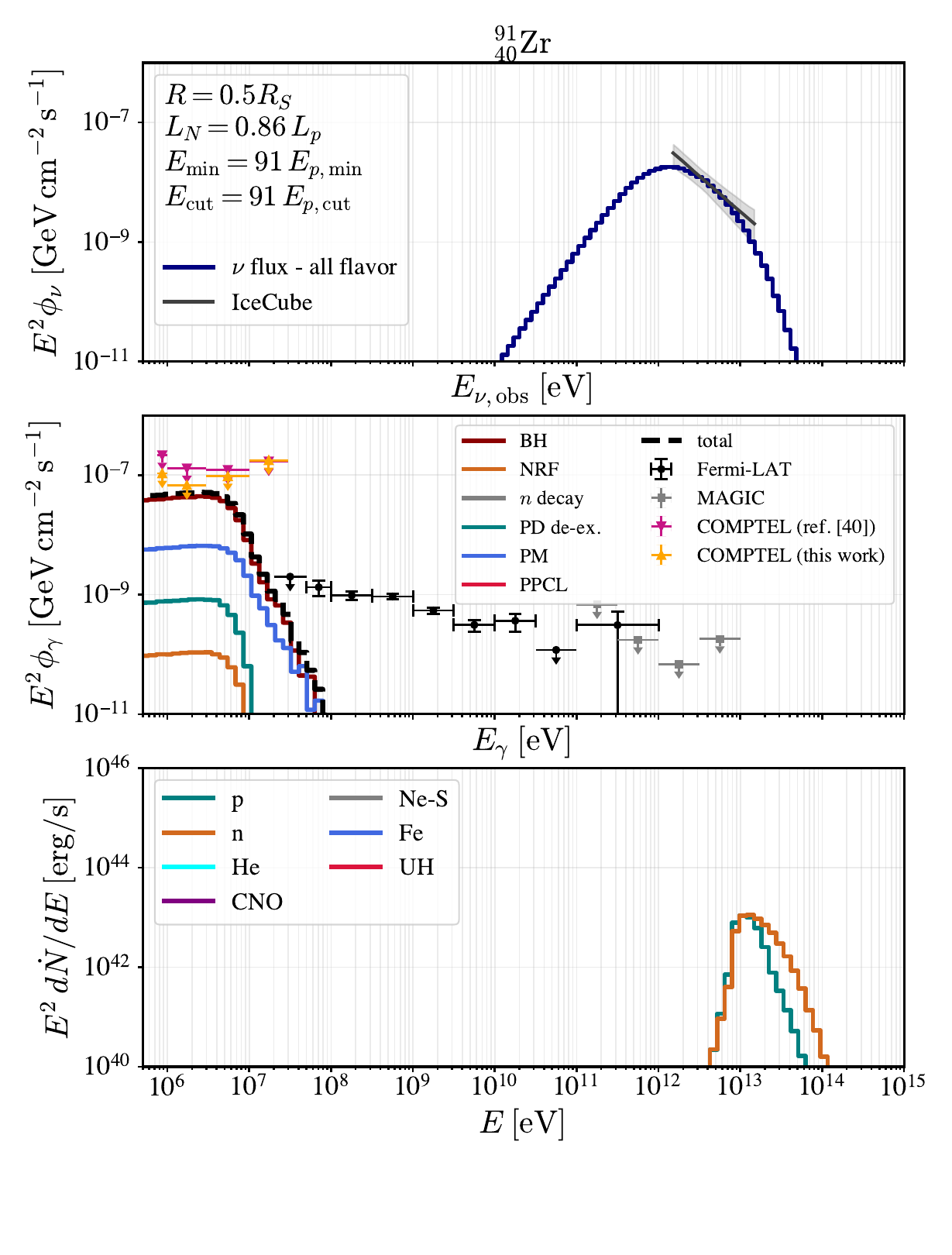}\label{fig:multimessenger_zirc_R0.5}}
\caption{Multimessenger spectra for the NGC~1068 model described in the text, for the compact/dense case $R=0.5\,R_S$. Panels in the left column present results for injecting protons, in the middle column for injecting Fe, in  the right column for injecting Zr. In each column, top panels show the neutrino fluxes, anchored to the IceCube inferred spectrum \cite{IceCube:2022der}, determining the overall normalization of the luminosities; middle panels show the gamma-ray flux, compared with  MAGIC data \cite{MAGIC:2019fvw} (gray), Fermi-LAT data \cite{Ajello:2023hkh} (black), existing COMPTEL bounds \cite{Schoenfelder:2000bu} (purple) and the ones recomputed in this work (yellow); bottom panels show the nuclear/nucleon content of the cosmic-ray flux, after crossing the region of size $R$.}
\label{fig:multimessenger_R0.5}
\end{figure*}

\begin{figure*}[tb]
\centering
\subfloat[]{\includegraphics[width=0.33\textwidth]{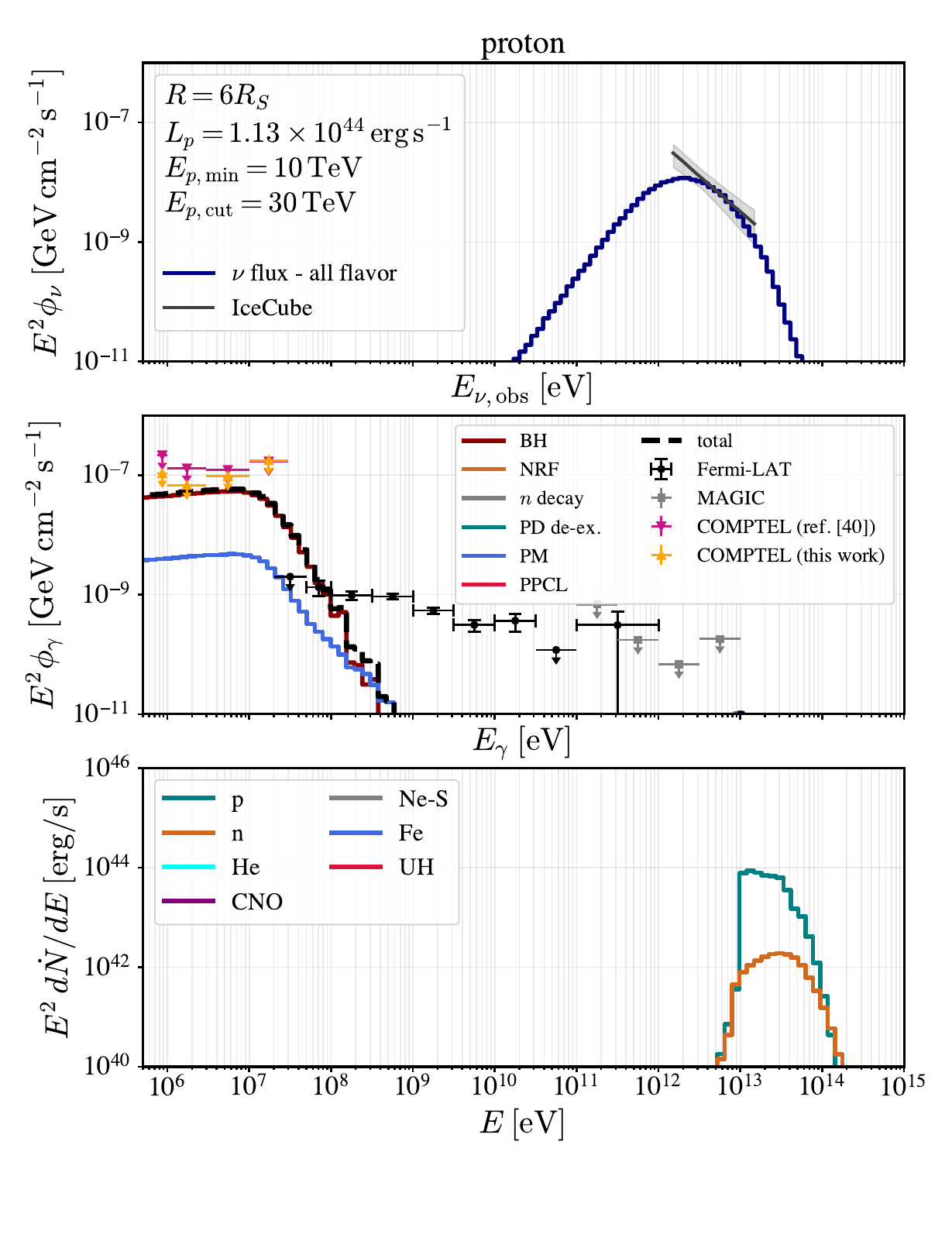}\label{fig:multimessenger_proton_R6}}
\hfill
\subfloat[]{\includegraphics[width=0.33\textwidth]{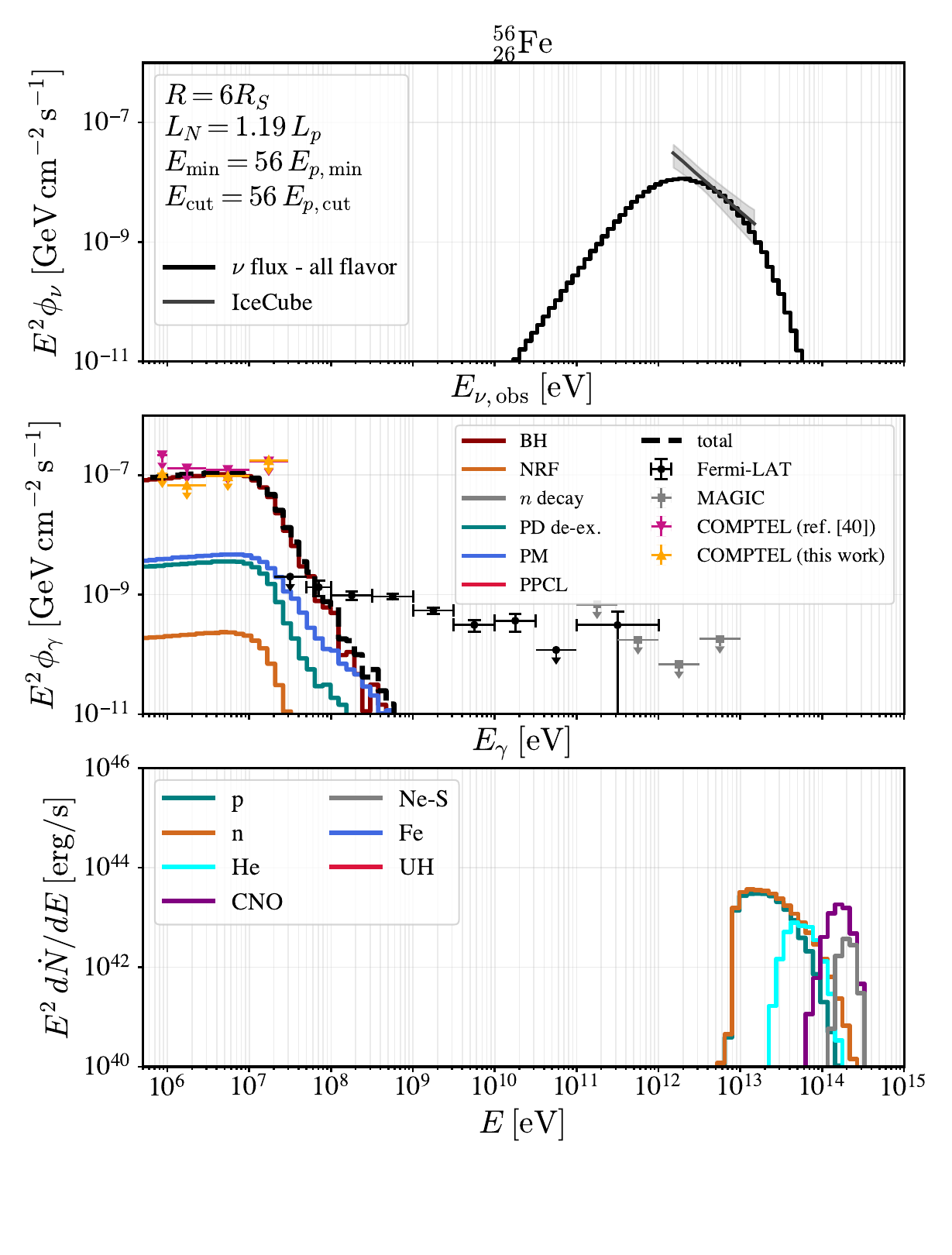}\label{fig:multimessenger_iron_6}}
\hfill
\subfloat[]{\includegraphics[width=0.33\textwidth]{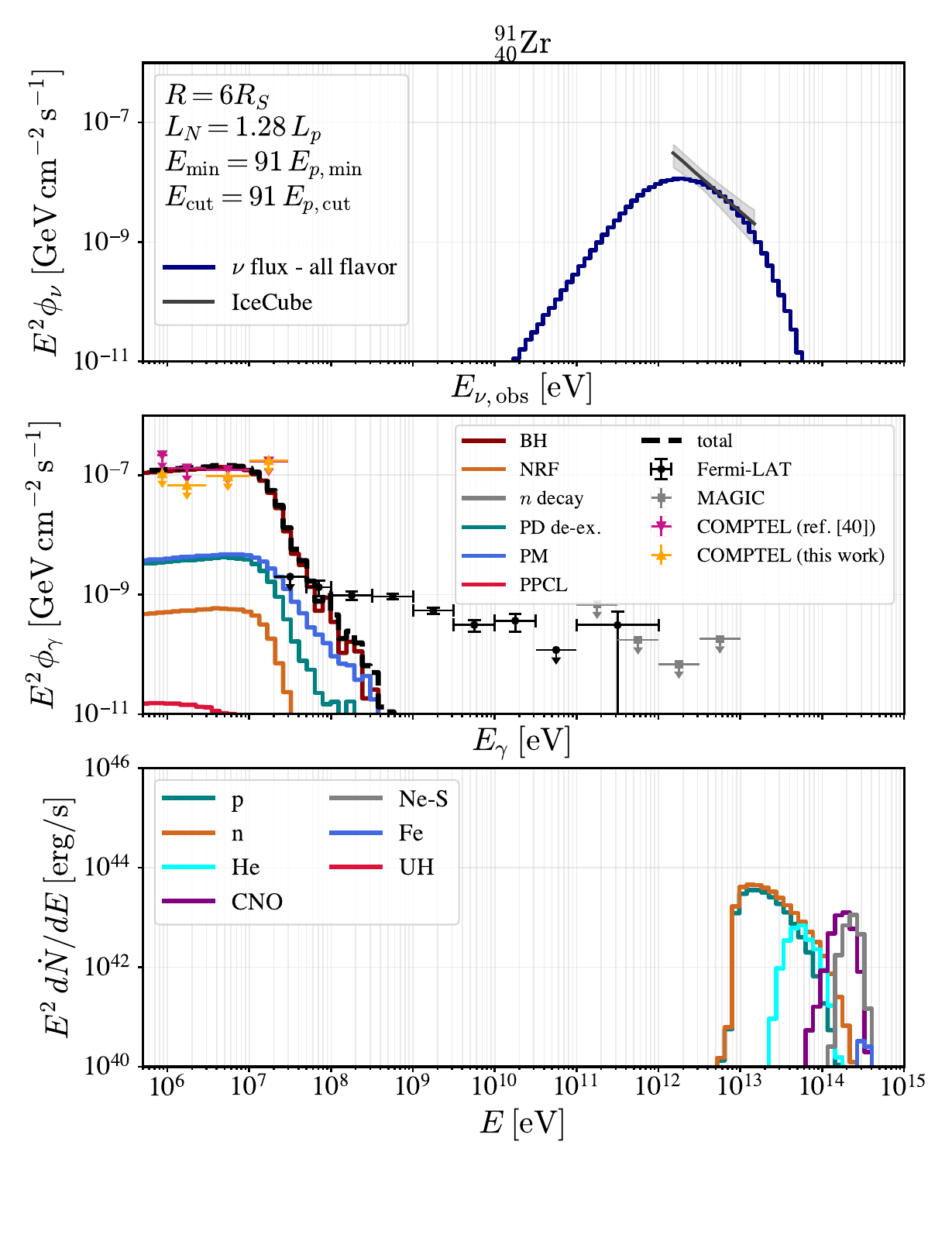}\label{fig:multimessenger_zirc_R6}}
\caption{As in Fig.~\ref{fig:multimessenger_R0.5}, for the more extended/rarefied case $R=6\,R_S$.}
\label{fig:multimessenger_R6}
\end{figure*}

Figures~\ref{fig:multimessenger_R0.5} and \ref{fig:multimessenger_R6} show how the change in the escaping composition translates into the full multimessenger output. Respectively, we display two representative cases: a compact, nucleon-dominated source with $R=0.5R_S$, and a more extended and diluted source with $R=6R_S$, where bound nuclear fragments still carry a substantial fraction ($\sim20\%-30\%$) of the escaping baryonic mass. In each figure, the upper panels show the neutrino flux after normalizing the model to the best-fit flux of NGC~1068 observed by IceCube; the middle panels show the gamma-ray emission after the electromagnetic cascade, separated by production channel; and the lower panels show the spectrum of nuclei escaping the emission region, grouped by mass.

Before discussing the different injected compositions, we first note that the proton injection reproduces the expected luminosity scale of the disk–corona model. After normalization to the IceCube flux, we find $L_p=3.24\times10^{43}\,{\rm erg\,s^{-1}}$ for $R=0.5R_S$, and $L_p=1.13\times10^{44}\,{\rm erg\,s^{-1}}$ for $R=6R_S$, which are consistent with the proton-injection results obtained in the same NGC~1068 disk-corona framework in ref.~\cite{Das:2024vug}. The luminosity factors quoted below for nuclear injection should therefore be interpreted relative to this proton baseline.

%%%%%%%%%%%%%%%%%%%%%%%%%%%%%%%
%%%%%%%%%%%% FIG. 10 %%%%%%%%%%%

%%%%%%%%%%%%%%%%%%%%%%%%%%%%%%%
%%%%%%%%%%%%%%%%%%%%%%%%%%%%%%%

%%%%%%%%%%%%%%%%%%%%%%%%%%%%%%%
%%%%%%%%%%%% FIG. 11 %%%%%%%%%%%

%%%%%%%%%%%%%%%%%%%%%%%%%%%%%%%
%%%%%%%%%%%%%%%%%%%%%%%%%%%%%%%

In the compact case, $R=0.5R_S$, the system lies in the nucleon-dominated regime identified above. Heavy nuclei are efficiently disintegrated before escape, and a sizable fraction of the baryonic energy is transferred to free neutrons. As discussed in Sec.~\ref{sec:multimessenger}, this reduces the amount of energy exposed to BH losses and enhances the fraction of the injected energy that can be converted into neutrinos. Consequently, for heavy nuclei injection the same IceCube neutrino normalization can be reached with a $\sim 15\%-20\%$ lower cosmic-ray luminosity than in the proton case. The accompanying electromagnetic cascade is also weaker,  since less energy is deposited into BH pairs: Integrating the cascaded photon output over energy, we find a bolometric gamma-ray luminosity lower by $\sim 35\%$ for Fe injection and by $\sim 28\%$ for Zr injection relative to the proton case. Thus, \textit{in a compact disk-corona scenario, heavy nuclei injection relaxes constraints from the MeV--GeV gamma-ray data, compared to pure proton injection}.

The situation is reversed for the more extended source, $R=6R_S$. In this case, photodisintegration does not develop completely over the propagation length, and a non-negligible fraction of the injected energy remains in charged nuclear fragments long enough or even after crossing a lenght $R$. These fragments suffer enhanced BH losses relative to protons. As a result, matching the IceCube flux requires a larger injected cosmic-ray luminosity for heavy nuclei. In our benchmark, the required luminosity is increased by about $20-30\%$ relative to proton injection. At the same time, the fraction of injected energy transferred to the electromagnetic cascade is also larger. The combination of a higher luminosity normalization and a larger electromagnetic energy budget enhances the cascaded gamma-ray flux by a factor of $\simeq2.0$ for Fe injection and $\simeq2.5$ for Zr injection relative to the proton benchmark, causing the heavy-nuclei scenarios to overshoot the MeV–GeV measurements and upper bounds. \textit{The gamma-ray constraints therefore become more restrictive for heavy-nucleus injection than for proton injection in this less compact regime.}

%%%%%%%%%%%%%%%%%%%%%%%%%%%%%%%%
%%%%%%%%%%%%%%%%%%%%%%%%%%%%%%%%
\section{The importance of the MeV gamma-ray observations}\label{subsec:comptel_ngc1068}
%%%%%%%%%%%%%%%%%%%%%%%%%%%%%%%%
%%%%%%%%%%%%%%%%%%%%%%%%%%%%%%%%

In our benchmark scenario, the source is highly opaque to GeV--TeV gamma rays, such that the electromagnetic cascade is efficiently reprocessed to lower energies. This makes the MeV observations particularly relevant since they can directly constrain the escaping low-energy cascade component. Motivated by this point, we performed a dedicated re-analysis of the archival COMPTEL observations of NGC~1068 using the modern \texttt{GammaLib}/\texttt{ctools} framework~\cite{Knodlseder:2016nnv}.

We consider a circular region of interest of radius \(10^\circ\) centered on the source position $(\mathrm{RA},\mathrm{Dec})=(40.6696^\circ,-0.0133^\circ)$,
and analyze the data in the four standard COMPTEL energy bands
\[
0.75\text{--}1~\mathrm{MeV},\quad
1\text{--}3~\mathrm{MeV},\quad
3\text{--}10~\mathrm{MeV},\quad
10\text{--}30~\mathrm{MeV}.
\]
For this selection, only the viewing period \texttt{vp0021\_0} contributes. In each band, NGC~1068 was modeled as a point source at fixed sky position and fitted independently.

No significant excess was found in any of the four bands. We therefore derived one-sided 95\% confidence level (C.L.) upper limits on the photon flux in each interval. We also tested the impact of adding explicit diffuse sky templates for Galactic bremsstrahlung and inverse-Compton emissions. In practice, the resulting limits changed only marginally, indicating that for the present setup, the upper limits are robust with respect to the diffuse template choice. In the following, we report the obtained results from the full fit including both the standard background treatment and the diffuse sky component.

The obtained upper limits are summarized in Table~\ref{tab:ngc1068_comptel_reanalysis}. Particularly, our re-analysis leads to stronger constraints compared with the previous COMPTEL analysis~\cite{Schoenfelder:2000bu}, improving the upper bounds by a factor of $\sim2$ in the first two energy bins. These results are shown in the middle panels of Figs.~\ref{fig:multimessenger_R0.5} and \ref{fig:multimessenger_R6}, where the upper limits derived in this work are depicted by yellow points while the bounds from ref.~\cite{Schoenfelder:2000bu}, over a quarter-century old, are shown by purple points for comparison.

%%%%%%%%%%%%%%%%%%%%%%%%%%%%%%%%%%
%%%%%%%%%%%% TABLE. I %%%%%%%%%%%% 
\begin{table}
\centering
\footnotesize
\caption{COMPTEL  95\% C.L. upper limits on NGC\,1068 from our re-analysis. The quoted differential limits are evaluated at the reference energy $E_{\rm ref}$ of each band, and $F_{95,\mathrm{band}}$ denotes the corresponding integrated photon-flux upper limit over that band.}
\label{tab:ngc1068_comptel_reanalysis}
\begin{tabular}{ccccc}
\hline
Energy band & \(E_{\rm ref}\) & TS &
\(\left.\mathrm{d}N/\mathrm{d}E\right|_{95}\) &
\(F_{95,\mathrm{band}}\) \\
(MeV) & (MeV) & &
(\(\mathrm{ph\,cm^{-2}\,s^{-1}\,MeV^{-1}}\)) &
(\(\mathrm{ph\,cm^{-2}\,s^{-1}}\)) \\
\hline
0.75--1.0   & 0.85 & 1.763 & \(1.43\times10^{-4}\) & \(3.56\times10^{-5}\) \\
1.0--3.0    & 2.0  & 1.413 & \(2.27\times10^{-5}\) & \(4.54\times10^{-5}\) \\
3.0--10.0   & 5.0  & 1.397 & \(3.19\times10^{-6}\) & \(2.23\times10^{-5}\) \\
10.0--30.0  & 15.0 & 0.201 & \(5.92\times10^{-7}\) & \(1.18\times10^{-5}\) \\
\hline
\end{tabular}
\end{table}

%%%%%%%%%%%%%%%%%%%%%%%%%%%%%%%%%%
%%%%%%%%%%%%%%%%%%%%%%%%%%%%%%%%%%

%%%%%%%%%%%%%%%%%%%%%%%%%%%%%%%%
%%%%%%%%%%%%%%%%%%%%%%%%%%%%%%%%
\section{Conclusions and Discussions}\label{sec:conclusion}
%%%%%%%%%%%%%%%%%%%%%%%%%%%%%%%%
%%%%%%%%%%%%%%%%%%%%%%%%%%%%%%%%

At the dawn of identifying astrophysical neutrino sources—with NGC 1068 and TXS 0506+056 already emerging as the first compelling candidates—we are entering the era of multimessenger astrophysics, in which combining different cosmic messengers enables access to source properties that are otherwise inaccessible through a single observational channel. One such property is the nuclear composition of the accelerated particles responsible for neutrino and gamma-ray production through hadronic interactions. In this work, we have investigated the imprints of source nuclear composition on the resulting neutrino and gamma-ray yields. Using the corona+disk model of NGC 1068 as a representative case, we have performed a Monte Carlo simulation of neutrino and gamma-ray production, incorporating the relevant photonuclear interaction channels and electromagnetic cascade processes within the source. In the cascade treatment, we have additionally implemented novel—although subdominant—processes such as muon pair production and electron pair production with capture into the $L$ shell.

By normalizing the neutrino yield to the IceCube observation of NGC 1068, we find that the source size plays a central role in determining how nuclear composition affects the multimessenger output. Compact sources, with characteristic sizes $R\sim R_S$, have large photonuclear depths and drive the nuclear cascade rapidly toward the free-nucleon regime. In this limit, heavy nuclei are efficiently fragmented before escape, and a sizable fraction of the baryonic energy is transferred to free neutrons. Since this neutral component avoids Bethe–Heitler losses before its first photomeson interaction, heavy-nucleus injection can yield neutrinos more efficiently than proton injection, while depositing less energy into the electromagnetic cascade. In our NGC~1068 benchmark, this allows the IceCube normalization to be reached with a smaller cosmic-ray luminosity and with a relaxed MeV–GeV gamma-ray output.

The situation changes in less compact regions. As $R$ increases, the photonuclear column depth decreases and photodisintegration becomes less complete over the source crossing time. The cascade then spends a larger fraction of its evolution in charged nuclear fragments rather than in free nucleons. These fragments suffer enhanced Bethe–Heitler losses relative to protons at the same energy per nucleon, while the delayed approach to the nucleon-dominated regime reduces the neutrino yield. Consequently, matching the IceCube flux requires a larger injected cosmic-ray luminosity for heavy nuclei, and the accompanying electromagnetic cascade is strengthened. In this regime, heavy-nucleus injection is therefore more tightly constrained by the MeV–GeV gamma-ray data than the corresponding proton scenario.

This makes the MeV band a particularly powerful diagnostic of hidden neutrino sources. In the disk–corona scenario considered here, the source is opaque to GeV–TeV photons, so the electromagnetic energy injected at high energies is reprocessed toward lower energies before escaping. Therefore, the resulting MeV emission is sensitive not only to the total cosmic-ray power, but also to the compactness of the emission region and the composition of the injected particles. Our COMPTEL re-analysis already provides more stringent upper limits that test this low-energy cascade component, and forthcoming MeV gamma-ray facilities, such as COSI \cite{Tomsick:2023aue} and proposed wide-field missions such as AMEGO-X \cite{Caputo:2022xpx} and e-ASTROGAM \cite{e-ASTROGAM:2016bph}, will be able to sharpen this probe substantially. 

The steps taken in this work represent an initial move toward unveiling the inner workings of astrophysical particle-production regions. More realism in the theoretical description of these sources may come from including synchrotron losses or 2D/3D geometry effects in the electromagnetic cascade, tasks that we plan to tackle in forthcoming works.  Additional environments, characteristic of different classes of sources, may be also tested. As more neutrino sources are identified and the power of the multimessenger approach continues to grow, we may soon be able to probe features of environments that have long remained hidden from direct observation. With neutrinos, gamma rays, and cosmic rays jointly illuminating the same sources from complementary perspectives, we move steadily toward a more complete picture of the complex interplay between high-energy particles and the extreme astrophysical engines that produce them: The multimessenger era has only just begun, and it will certainly clarify many mysteries the Universe has yet to reveal.

%%%%%%%%%%%%%%%%%%%%%%%%%%%
\begin{acknowledgments}
%%%%%%%%%%%%%%%%%%%%%%%%%%%
A.F. E. acknowledges support from the São Paulo Research Foundation (FAPESP) through a postdoctoral fellowship, Grant No. 2025/03199-0. P.D.S. acknowledges support by Université Savoie Mont Blanc (USMB) via the TURBCOS grant. 
%%%%%%%%%%%%%%%%%%%%%%%%%%%
\end{acknowledgments}
%%%%%%%%%%%%%%%%%%%%%%%%%%%

\bibliography{refs}

\end{document}